\documentclass[11pt,preprint,aps,showkeys,onecolumn]{revtex4}
\usepackage{mathrsfs}
\usepackage{amsmath}
\usepackage{amssymb}
\usepackage{amsfonts}
\usepackage{tikz}
\usepackage{xcolor}
\usepackage{epstopdf}
\usepackage{dcolumn}
\usepackage{bm}
\usepackage{subcaption}
\usepackage{natbib}
\usepackage{float}
\usepackage{afterpage}
\usepackage{mathtools}
\usepackage{graphicx}
\usepackage[utf8]{inputenc} 
\usepackage[T1]{fontenc}
\usetikzlibrary{decorations.pathmorphing}
\usepackage[section]{placeins}

\usepackage{hyperref}
\begin{document}

\title{Absorption and scattering of massless scalar waves by Frolov black holes}
\author{Jining Tang$^{1}$}
\email{aishiker1998@gmail.com}
\author{Yang Huang$^{2}$}
\email{sps\_Huangy@ujn.edu.cn}
\author{Hongsheng Zhang$^{2}$}
\email{sps\_zhanghs$@$ujn.edu.cn (corresponding author)}
\affiliation{School of Physics and Technology, University of Jinan,
336 West Road of Nan Xinzhuang, Jinan, Shandong 250022, China}

\begin{abstract}
We study the absorption and scattering of massless scalar waves by Frolov black holes, a regular deformation of the Reissner--Nordstr\"om geometry. A null-geodesic analysis provides the photon-sphere radius and the critical impact parameters governing capture and glory scattering. Using a partial-wave approach, we compute total/partial absorption cross sections and differential scattering cross sections over a broad frequency range. When the absorption spectrum is rescaled by photon-sphere scales, the high-frequency oscillations exhibit a pronounced data collapse in the variables $\hat{\sigma}=\sigma_{\rm abs}/\sigma_{\rm geo}$ and $x=\omega/\Omega_c$, highlighting photon-sphere control of the absorption fine structure. We clarify the parameter dependence under the horizon-radius normalization and relate the apparent trends to the variation of the dimensionless mass across parameter space. Finally, comparing Frolov, Reissner--Nordstr\"om, and Hayward black holes with matched critical or glory impact parameters, we find that their absorption and scattering patterns can become remarkably close, indicating that in the intermediate-to-high frequency regime the dominant signatures are largely governed by the unstable photon orbit, while core effects enter as subleading corrections.
\end{abstract}
\maketitle	
\section{Introduction}
 
With the advent of gravitational-wave astronomy and black hole imaging\cite{maggiore2008gravitational1,maggiore2008gravitational2,creighton2012gravitational,sathyaprakash2009physics,hughes2003listening,abbott2016observation}, understanding the interaction between waves and black holes has become more crucial than ever. The scattering and absorption of fields by black holes provide powerful probes of spacetime structure and tests of general relativity\cite{anderson2002scattering,newton2013scattering}. Related wave-scattering phenomena have also been discussed in alternative gravitational settings, e.g., for (quasi-)spherical gravitational waves scattered from black strings in massive gravity\cite{zhang2021spherical}. Propagating fields leave characteristic imprints in their absorption and differential scattering cross sections, which depend both on universal near-horizon physics and on the detailed geometry of the photon sphere and its unstable null orbits. At low frequencies these observables typically display universal behaviour, largely insensitive to the fine details of the spacetime \cite{das1997universality}, whereas at high frequencies they are controlled by the capture properties of null geodesics and by the structure of the photon sphere, which also underlies the phenomenology of black hole shadows and photon rings as observed in M87* and Sgr~A* \cite{EHT2019I,EHT2022I}. Wave-optics diffraction phenomena can also arise in black-hole scattering, such as the Poisson–Arago spot discussed for gravitational waves\cite{zhang2021poisson}.

For a minimally coupled massless scalar field, this picture is well understood in benchmark geometries. In the zero-frequency limit the total absorption cross section approaches the area of the event horizon \cite{higuchi2001low,higuchi2002addendum}, while in the high-frequency regime it oscillates around the geometric capture cross section, with an oscillatory pattern that can be related to the properties of the unstable photon orbit and the associated Regge poles of the scattering matrix \cite{decanini2011universality,sanchez1978absorption}. These features have been established in detail for Schwarzschild and Reissner--Nordstr\"om black holes within the partial-wave framework pioneered for black-hole scattering by S\'anchez \cite{sanchez1978elastic} and developed further in subsequent work (see, e.g., \cite{FHMscattering,crispino2009scattering}). Full-wave computations of scalar scattering in nontrivial charged backgrounds,
together with comparisons to geodesic and Regge-pole/glory approximations,
have been carried out for charged dilatonic black holes \cite{huang2020scattering}.

The quest to resolve the central singularities predicted by general relativity has motivated the study of regular black holes (RBHs) \cite{ansoldi2008spherical,lan2023regular,bronnikov2007regular,bambi2023regular}. Many such solutions arise in nonlinear electrodynamics and modify the inner structure while preserving asymptotic flatness \cite{ayon1998regular,ayon2000bardeen,balart2014regular}. In particular, the Bardeen and Hayward spacetimes \cite{bardeen1968non,hayward2006formation} provide simple models in which regular cores and effective charges reshape the effective potential and photon-sphere data, leading to characteristic changes in greybody factors and large-angle scattering patterns \cite{macedo2014absorption,macedo2015scattering}.

As a charged generalization of the Hayward solution, the Frolov black hole \cite{frolov2016notes} has recently attracted considerable interest. Its thermodynamic properties, quasinormal modes and shadows have been investigated in a variety of contexts \cite{murk2023regular,song2024quasinormal,Kala:2025FrolovRBH}. However, to the best of our knowledge, a systematic study of massless-scalar absorption and scattering in this spacetime, covering analytic low- and high-frequency limits and full-frequency numerical results, is still lacking.

Against this backdrop, the present work develops a comprehensive analysis of massless scalar absorption and scattering by Frolov black holes. At the classical level we study null geodesics, identifying the photon sphere and the associated critical impact parameters, and we derive the weak-deflection and backward glory approximations for the differential cross section. Within the partial-wave formalism we obtain the radial equation and its effective potential, and we derive the usual low-frequency limit and a high-frequency sinc approximation expressed in terms of the critical impact parameter and Lyapunov exponent. We then compute the absorption and differential scattering cross sections numerically over a broad frequency range, comparing them with the analytic approximations and exploring how the two Frolov parameters jointly affect the spectra. In particular, we examine cases in which different parameter choices yield very similar absorption and scattering patterns, and relate this behaviour to the structure of the effective potential and the underlying geodesics.

The remainder of this paper is organized as follows. In Sec.~\ref{sec:classical} we review the Frolov spacetime, analyse null geodesics and discuss the corresponding classical capture and scattering properties. In Sec.~\ref{sec:partial} we present the partial-wave formalism for a massless scalar field, derive the effective potential, and obtain the analytic low- and high-frequency limits for the absorption and scattering cross sections. Sec.~\ref{sec:numerics} is devoted to the numerical results: we summarize our normalization choices, describe the numerical implementation, and present the full-frequency absorption and scattering spectra for a range of Frolov parameters, including comparisons with Reissner--Nordstr\"om and Hayward black holes at matched impact parameters. We conclude our results in Sec.~\ref{sec:conclusion}. Throughout this work we use natural units with $G=c=\hbar=1$ and metric signature $(-,+,+,+)$. For analytic expressions we keep the mass parameter $M$ explicit to facilitate comparison with the existing literature, whereas in our numerical calculations and plots we adopt the horizon-radius normalization $r_h=1$, defined by $f(r_h)=0$.

\section{Classical Analysis}
\label{sec:classical}

We consider the Frolov BH, which is a static regular black hole, whose line element is given by \cite{frolov2016notes}
\begin{equation}
    d s^2=-f(r)d t^2+\frac{1}{f(r)}d r^2+r^2\left(d \theta^2+\sin^2\theta d \phi^2 \right),
\end{equation}
where 
\begin{equation}
    f(r)= 1 - \frac{r^2 (2Mr-Q^2)}{r^4+(2Mr+Q^2)\alpha^2}.\label{eq:metric}
\end{equation}
Here $M$ and $Q$ respectively denote the mass and charge parameter of the black hole. The regularized parameter $\alpha$ represents the Hubble length, which is related to the effective cosmological constant $\Lambda_{\text{eff}}=3/\alpha^2$, characterize the curvature core of the Frolov BH. For keeping the event horizon surviving, the Hubble length is bounded with $\alpha\leq\sqrt{16/27}M$. 

As $r\to \infty$, the spacetime is asymptotically Reissner--Nordstr\"om $f(r)\approx f_{\text{RN}}(r)=1-\frac{2M}{r}+\frac{Q^2}{r^2}$, where near the center $r\to 0$ the expansion $f(r)= 1-\frac{\Lambda_{\text{eff}}}{3}r^2 +\cdots$ signals regularity.

\subsection{Null geodesics and geometrical limits}
\label{subsec:geo}
The classical Lagrangian related to the null geodesic trajectory is given by  

\begin{equation}
    \mathcal{L}=\frac{1}{2}g_{\mu\nu}\frac{d x^\mu}{d \lambda}\frac{d x^\nu}{d \lambda}=0.
\end{equation}
Let us constrain the motion of particles on the equatorial plane without loss of generality, the equation above gives 
\begin{equation}
    f(r)\dot{t}^2-\frac{\dot{r}^2}{f(r)}-r^2\dot{\phi}^2=0.
\end{equation}
It's clear that there exist two conserved quantities 
\begin{equation}
    E=\frac{\partial\mathcal{L}}{\partial\dot{t}}=-f\dot{t}\quad,\quad L=-\frac{\partial\mathcal{L}}{\partial\dot{\phi}}=r^2\dot{\phi},
\end{equation}
where $E$ and $L$ are respectively the energy and the angular momentum of the massless particle.
Then the geodesic equation can be written as 
\begin{equation}
    \dot{r}^2=f^2 \dot{t}^2-fr\dot{\phi}^2=E^2-f\frac{L^2}{r^2},
\end{equation}
here we can write the equation of motion Eq.(6) in the form 
\begin{equation}
    \dot{r}^2+V_{\text{eff}} = E^2
\end{equation}
by defining 
\begin{equation}
    V_{\text{eff}} =f(r)\frac{L^2}{r^2} =\left[1 - \frac{r^2 (2Mr-Q^2)}{r^4+(2Mr+Q^2)\alpha^2}\right]\frac{L^2}{r^2}
\end{equation}
In Fig.\ref{fig:veff}, we display the effective potential for massless particles in Frolov BHs with different parameters for illustration.

After using the impact parameter $b\equiv L/E$ as the ratio of angular momentum and energy, we can obtain the orbit equation 
\begin{equation}
    \left(\frac{d r}{d \phi}\right)^2=\frac{\dot{r}^2}{\dot{\phi}^2}=r^4\left(\frac{1}{b^2}-\frac{f}{r^2}\right).\label{oeq}
\end{equation}
It is convenient to introduce  $u\equiv 1/r$, the orbit equation can be written as 
\begin{equation}
    \left(\frac{d u}{d \phi}\right)^2=\left(\frac{d u}{d r}\right)^2\left(\frac{d r}{d \phi}\right)^2=\mathcal{U}(u)=\frac{1}{b^2}-u^2 f,
    \label{eq:1order}
\end{equation}
where $f=f(1/u)$.
By differentiating Eq.\eqref{eq:1order}, we get
\begin{equation}
    \frac{d ^2 u}{d \phi^2}=-\frac{u^2}{2}\frac{df}{du}-u f
\end{equation}
For the Frolov spacetime, 
\begin{equation}
    f(r)=1-\frac{(2Mr-Q^2)r^2}{r^4+(2Mr+Q^2)\alpha^2},
\end{equation}
two orbit equations read
\begin{equation}
    \left(\frac{d u}{d \phi}\right)^2=\frac{1}{b^2}-u^2+\frac{u^3 \left(2 M-Q^2 u\right)}{\alpha^2 u^3 \left(2 M+Q^2 u\right)+1},
\end{equation}

\begin{equation}
\frac{d^{2}u}{d\phi^{2}} \;=\; -\,\frac{P(u)}{B(u)^{2}}, 
\label{eq:u-phi-ode}
\end{equation}
where
\begin{equation}
    B(u) = 1+\alpha^{2}u^{3}\bigl(2M+Q^{2}u\bigr),
\end{equation}
\begin{equation}
\begin{aligned}
P(u) = {}& u\Bigl[\,1 - 3Mu + 2Q^{2}u^{2}
+ 4\alpha^{2}Mu^{3} + 2\alpha^{2}Q^{2}u^{4} + 2\alpha^{2}M Q^{2}u^{5} \\
&\hspace{5.5em}
+\, \alpha^{4}\bigl(4M^{2}u^{6}+4MQ^{2}u^{7}+Q^{4}u^{8}\bigr)\Bigr].
\end{aligned}
\label{eq:P-of-u}
\end{equation}

It is obvious that, for the curvature-free case ($\alpha=0$),
\begin{equation}
    f=1- \frac{2M}{r}+\frac{Q^2}{r^2}=1-2Mu+Q^2u^2,
\end{equation}
the equations are consistent with previous work on RN BHs \cite{crispino2009scattering,chandrasekhar1998mathematical}
\begin{equation}
    \left(\frac{d u}{d \phi}\right)^2=\frac{1}{b^2}-u^2+2Mu^3-Q^2u^4,
\end{equation}
\begin{equation}
     \frac{d ^2 u}{d \phi^2}=-u+3Mu^2-2Q^2u^3.
\end{equation}

\begin{figure}[htb]
    \centering 
    \begin{subfigure}[b]{0.48\textwidth}
        \centering
        \includegraphics[width=\textwidth]{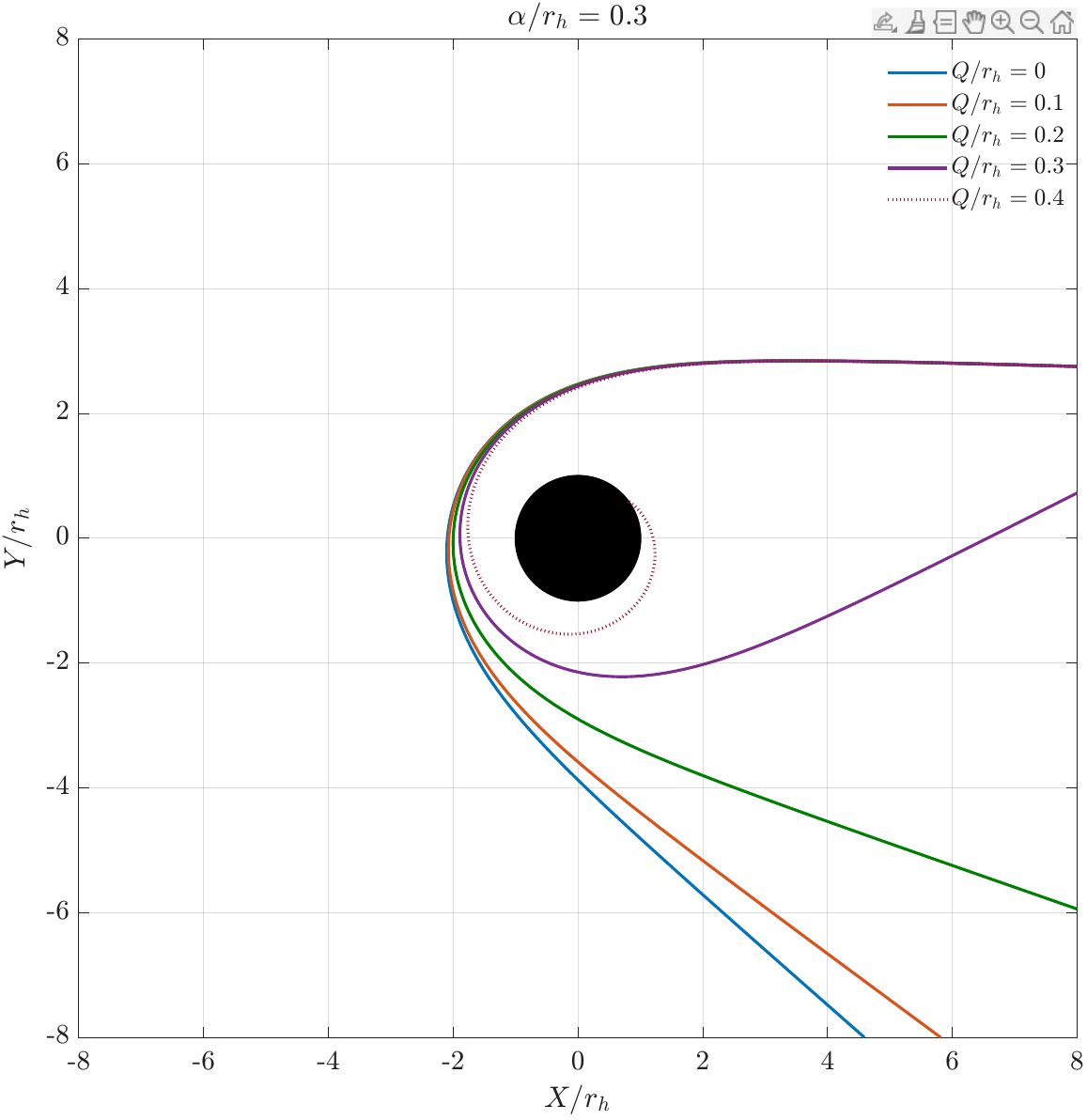} 
        \caption{Null geodesics for different $Q/r_h$ with $\alpha/r_h = 0.3$}
        \label{fig:geoQ}
    \end{subfigure}
    \begin{subfigure}[b]{0.48\textwidth}
        \centering
        \includegraphics[width=\textwidth]{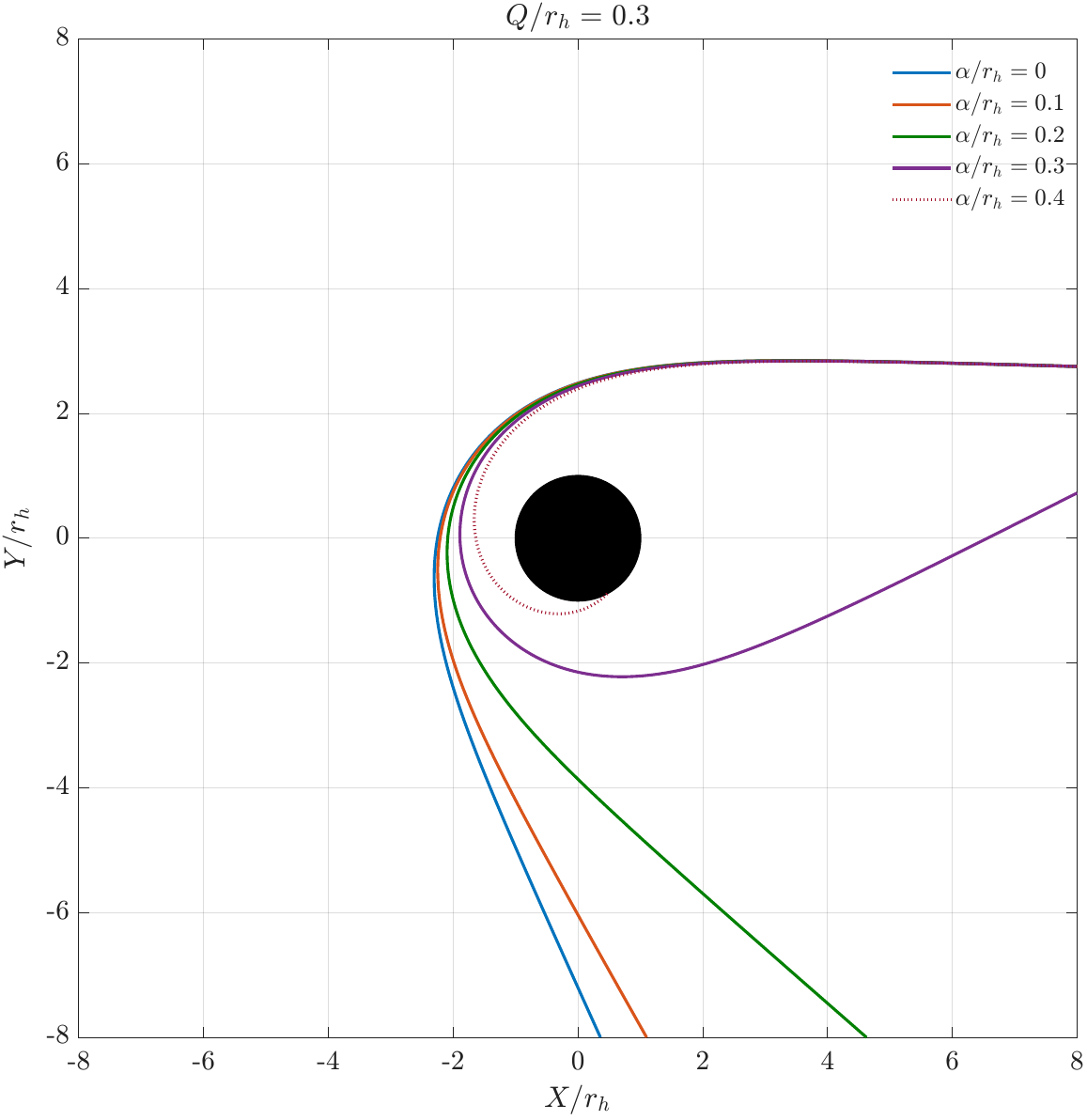} 
        \caption{Null geodesics for different $\alpha/r_h$ with $Q/r_h = 0.3$}
        \label{fig:geoL} 
    \end{subfigure}
    \caption{
         Null geodesics for massless particles in the background of Frolov BHs with different values of $Q$ and $\alpha$ for a fixed $b=3 r_h$. The trajectories can be derived numerically by orbit equation Eq. \eqref{eq:1order}. We have set the initial condition $r_{\rm inf}$ at $100 r_h$.
    }
    \label{fig:geodesic} 
\end{figure}

From the orbit equation Eq.\eqref{eq:1order}, one sees $\mathcal{U}(u)\geq 0$ along the orbit. There is a critical value of the impact parameter $b_c$, which related to the critical radius $r_c$ of the unstable circular orbit, namely
\begin{equation}
    b_c=\frac{L_c}{E_c}=\frac{r_c}{\sqrt{f(r_c)}},\quad 2f(r_c)-r_c\frac{df(r_c)}{dr}=0.  \label{eq:phsph}
\end{equation}
The trajectories can be obtained by numerically solving the orbit equation \eqref{oeq}. In Fig.\ref{fig:geodesic}, we give the null geodesic orbits in Frolov spacetime with several parameters. In particular, the photon-sphere data entering $b_c$ also control the CAM ingredients, like Regge frequency and damping, as explicitly demonstrated in Ref.~\cite{li2022absorption}.
 
 These null-geodesic quantities, in particular the photon-sphere radius and the critical impact parameter $b_c$, also underlie the optical appearance of compact objects.
Recent analyses of photon dynamics and shadows in rotating spacetimes with parity-odd hair provide
closely related geodesic diagnostics \cite{huang2025dynamics}. See also Ref.~\cite{meng2023dynamics} for a systematic discussion of null-particle dynamics and shadows
in a general rotating black-hole geometry. Light rings and lensing diagnostics are not restricted to black holes; for related discussions in rotating boson-star spacetimes
with parity-odd features, see Ref.~\cite{huang2025lensing}.

 Equation \eqref{eq:phsph} partitions incoming null geodesics into two classes:
for $|b|<b_c$ the trajectory crosses the potential barrier and is
captured, whereas for $|b|>b_c$ it turns back to infinity. The geometric cross section, which is the classical capture cross section of geodesics, is well known\cite{Wald:1984}
\begin{equation}
    \sigma_{\rm geo}=\pi b_c^2.
\end{equation}

In high frequency limit, the absorption cross section can be written as the sinc approximation, which involves leading term $\sigma_{\rm geo}$ and the oscillatory correction\cite{sanchez1978absorption,decanini2011universality}
\begin{equation}
    \sigma_{\rm abs}^{\rm hf}\approx \sigma_{\rm geo}\left[1-8\pi \frac{\Lambda_c}{\Omega_c}  e^{-\pi \Lambda_c/\Omega_c}\operatorname{sinc} (2\pi b_c \omega)\right],
\end{equation}
where $\operatorname{sinc}(x)\equiv\sin(x)/x$. 
And the photon-sphere angular frequency and Lyapunov exponent\cite{cardoso2009geodesic}, $\Omega_c=1/b_c$ is the angular frequency of the unstable photon circular orbit at $r_c$. Lyapunov exponent of these circular null geodesics is given by\cite{cardoso2009geodesic}
\begin{equation}
\Omega_c=\frac{1}{b_c}=\frac{\sqrt{f(r_c)}}{r_c},\quad
\Lambda_c=\sqrt{\left.\frac{L^2}{2 \dot{t}^2}V''(r)\right|_{r_c}}=\sqrt{\frac{f(r_c)\Bigl[2f(r_c)-r_c^2 f''(r_c)\Bigr]}{2r_c^2}},
\end{equation}
will anchor the oscillatory pattern of the high-frequency absorption in Sec.~\ref{sec:absorption}.

\subsection{Weak deflection and Classical scattering}

Having established the geometrical-optics limit via the photon sphere
(\(r_c,b_c\)) and the capture cross section \(\sigma_{\rm geo}\) in the previous subsection, we now focus on the rays that return to infinity, i.e. the scattering branch $b>b_c$. 
In the scattering case, the deflection angle can be obtained by integrating Eq.\eqref{eq:1order}
\begin{equation}
    \Theta(b)=2\int_0^{u_0}\frac{du}{\sqrt{\mathcal{U}(u)}}-\pi=2\int_0^{u_0}du\left(\frac{1}{b^2}-u^2 f\right)^{-1/2}-\pi.
\end{equation}
In the weak-field limit, the deflection angle can be expanded in power of $b^{-1}$
\begin{equation}
    \Theta\approx\frac{4M}{b}+\frac{3\pi}{4b^2}\left(5M^2-Q^2\right)+\mathcal{O}\left(\frac{1}{b}\right)^3 ,
    \label{eq:deflect}
\end{equation}
which coincides with the RN result. Expanding the lapse at large radii yields
\begin{equation}
f(r)=1-\frac{2M}{r}+\frac{Q^{2}}{r^{2}}
+\alpha^{2}\!\left(\frac{4M^{2}}{r^{4}}-\frac{Q^{4}}{r^{6}}\right)+\cdots,\label{eq:expand}
\end{equation}
so the $\alpha$-dependent corrections start at $O(r^{-4})$ and therefore enter the
weak-deflection series only at $O(b^{-4})$. This is consistent with the weak-deflection expansion obtained via the Gauss-Bonnet approach in Ref.~\cite{Kala:2025FrolovRBH} and with the Hayward case~\cite{de2023geodesic}.

The classical differential cross section is given by  \cite{newton2013scattering}

\begin{equation}
\left.\frac{d\sigma_{\rm sc}}{d\Omega}\right|_{\rm cl}
= \frac{b(\Theta)}{\sin\theta}\left|\frac{db(\Theta)}{d\Theta}\right|,
\label{eq:classical-dsigma}
\end{equation}
where $\theta$ is the scattering angle, and $\Theta$ is the deflection angle.
From Eq.\eqref{eq:deflect} and \eqref{eq:classical-dsigma}, we conclude that the classical differential scattering cross section for small scattering angles is
\begin{equation}
    \left.\frac{d\sigma_{\rm sc}}{d\Omega}\right|_{\rm cl}=\frac{16M^2}{\Theta^4}+\frac{3\pi (5M^2-Q^2)}{4\Theta^3}+\mathcal{O}(\Theta^{-2}) .
    \label{eq:classical dcs}
\end{equation}
Eq.~\eqref{eq:classical dcs} immediately connects with known results for other black-hole spacetimes. For $Q=0$ it reduces to the Schwarzschild small-angle cross section, whose universal leading term $16M^2/\Theta^{4}$ corresponds to gravitational Rutherford scattering, together with the standard correction $\propto \Theta^{-3}$ \cite{sanchez1978elastic,newton2013scattering}.
For $\alpha=0$ the expression coincides with the Reissner--Nordstr\"om case, where the electric charge first enters at order $\Theta^{-3}$ and decreases the cross section relative to Schwarzschild \cite{crispino2009scattering}. In the Frolov spacetime the parameter $\alpha$ does not contribute up to order $\Theta^{-3}$, this follows from the large-$r$ expansion of $f(r)$(see Eq.~\eqref{eq:expand}), where the $\alpha$-dependent terms start at $\mathcal{O}(r^{-4})$, so that they affect the weak-deflection series only at higher orders, as also observed in the Hayward case \cite{de2023geodesic}.

For the glory approximation of the scalar scattering cross sections, the backward glory is defined by
$\Theta(b_g)=\pi$, in the spacetime of a static and symmetric BH, the glory approximation of the scalar scattering cross sections is \cite{matzner1985glory}
\begin{equation}
    \left.\frac{d \sigma_{s c}}{d \Omega}\right|_{\theta \approx \pi} \approx 2 \pi \omega b_g^2\left|\frac{d b}{d \theta}\right|_{\theta=\pi}\left[J_0\left(\omega b_g \sin \theta\right)\right]^2 ,\label{eq:glory}
\end{equation}
where $\omega$ is the frequency of the scalar wave, $b_g$ is the impact parameter that corresponds to a deflection angle of $\pi$, $J_0(x)$ is a Bessel function of the first kind. Imposing the backward-scattering condition $\theta=\pi+2 n\pi $ ($n=0,1,2,\ldots$) yields a discrete set of glory impact parameters $b_g^n$, where $n$ counts the number of windings of the null geodesic around the photon sphere. In practice, the differential cross section near $\theta\simeq\pi$ is dominated by the $n=0$ branch, while higher-winding contributions are strongly suppressed \cite{macedo2015scattering,de2022scattering}. Although the semi-classical glory approximation~\eqref{eq:glory} is formally derived for the high-frequency regime, it remains in very good agreement with the numerical results down to intermediate frequencies \cite{crispino2009scattering}.

\section{Partial-Wave Method}
\label{sec:partial}
\subsection{Massless scalar field}
First we consider a minimally coupled massless scalar obeying the Klein-Gordon equation
\begin{equation}
    \nabla_\mu\nabla^\mu\Phi=\frac{1}{\sqrt{-g}}\partial_\mu\left(\sqrt{-g}g^{\mu\nu}\partial_\nu \Phi\right)=0.
    \label{eq:KG}
\end{equation}
With the standard separation
\begin{equation}
    \Phi=\frac{\psi_{\omega l}(r)}{r}Y_{lm}(\theta,\phi)e^{-i\omega t},
    \label{eq:decompose}
\end{equation}
the radial function satisfies a RW-type equation in the tortoise coordinate
\begin{equation}
    \frac{d^2}{dr_*^2}\psi_{\omega l}+\left[\omega^2-V_{\rm eff}(r)\right]\psi_{\omega l}=0,
    \label{eq:radial}
\end{equation}
with effective potential
\begin{equation}
    V_{\rm eff}(r)=f(r)\left[\frac{f'(r)}{r}+\frac{l(l+1)}{r^2}\right],
    \label{eq:veff}
\end{equation}
where tortoise coordinate $r_*$ is defined as $f(r)dr_*=dr$.
\begin{figure}[H] 
    \centering 
    \includegraphics[width=0.8\textwidth]{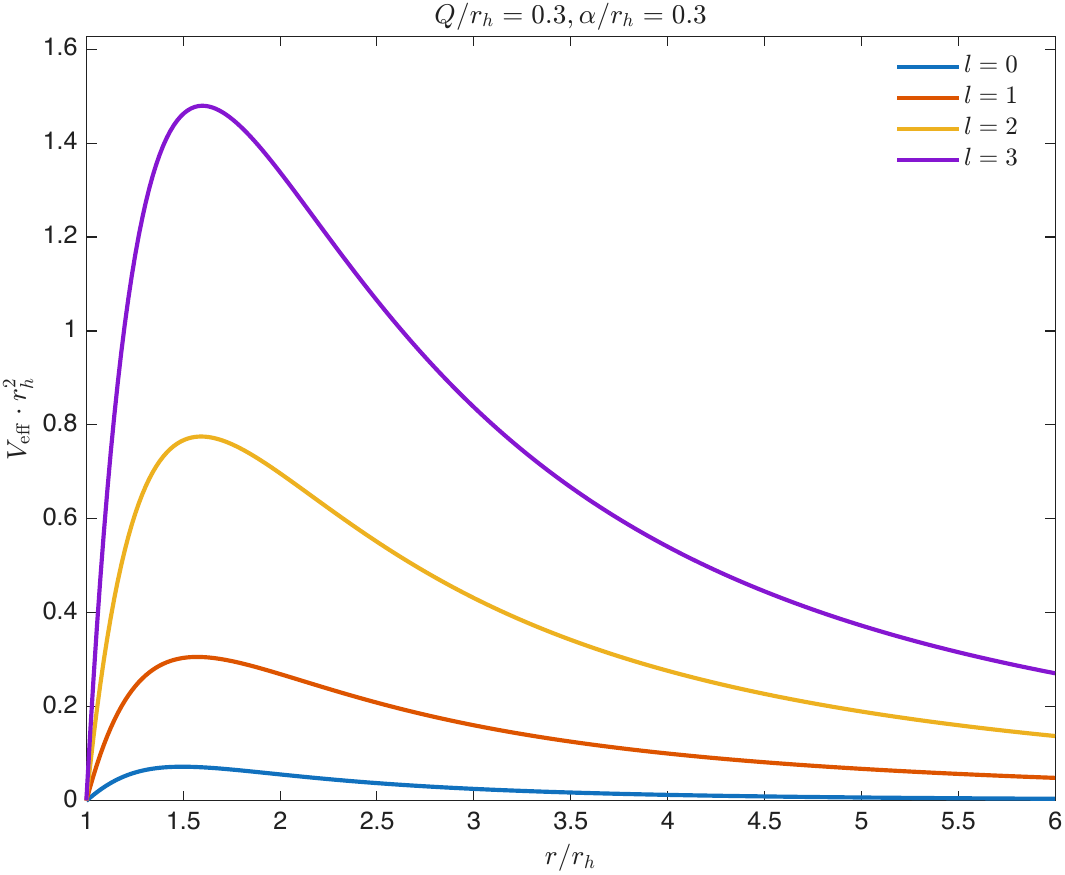}
    \caption{
        The effective potential of a Frolov BH with the parameters $Q/r_h=0.3$ and $ \alpha/r_h = 0.3$.
    }
    \label{fig:veff} 
\end{figure}

\begin{figure}[H] 
    \centering 
    \includegraphics[width=0.8\textwidth]{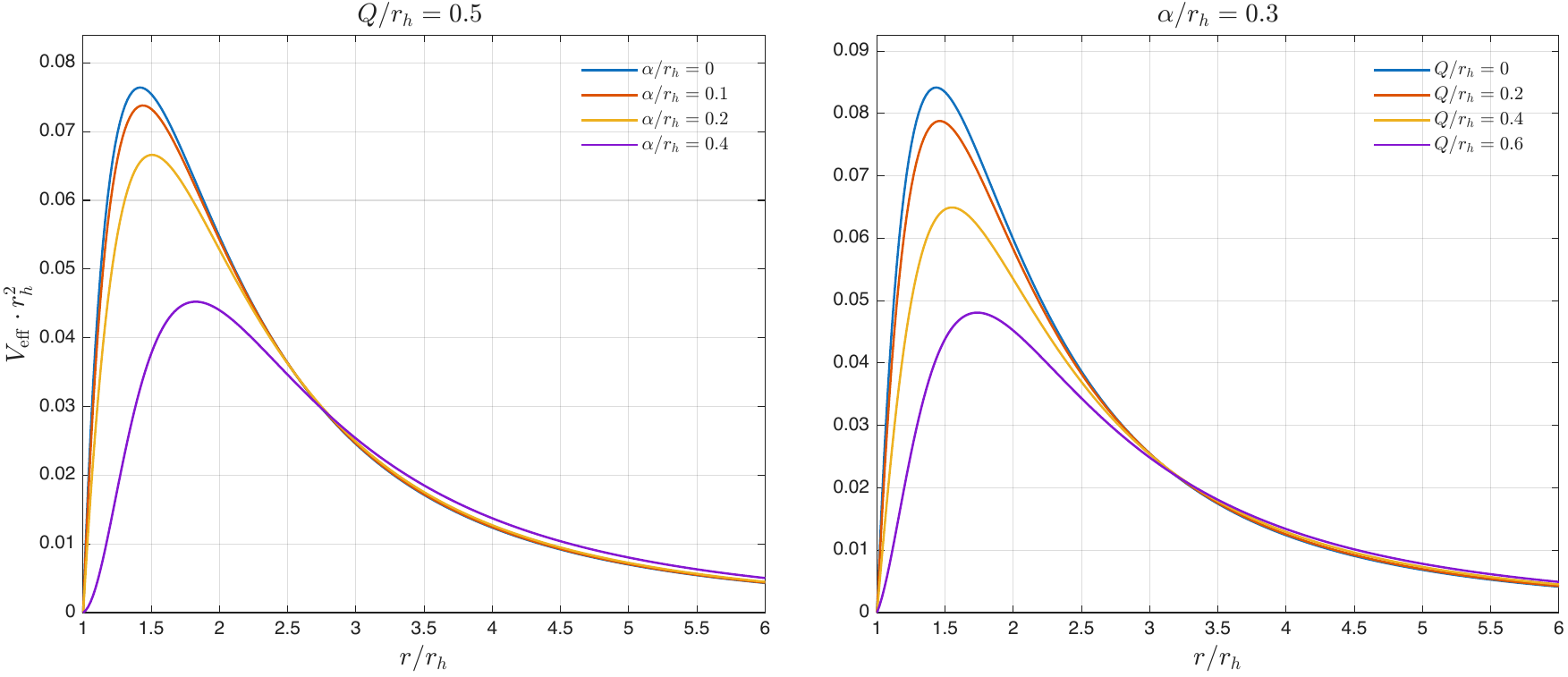}
    \caption{
        Effective potential $V_{\text{eff}}(r)$ for the $l=0$ mode of Frolov BHs.
    }
    \label{fig:veffpara} 
\end{figure}

In Fig.~\ref{fig:veff} we plot the effective potential for a "mild" Frolov black hole with parameters chosen as $Q/r_h=0.3$ and $\alpha/r_h=0.3$.
Fig.~\ref{fig:veffpara} compares the effective potential, with $l=0$ for various parameters. As we can see, no matter which parameter is fixed, the maximum of the potential always decreases as the other parameter increases. This apparent discrepancy with the trend reported in works that adopt the $M=1$ normalization for RN and Hayward black holes arises solely from the different choice of units; a detailed discussion of the relation between the $r_h=1$ and $M=1$ conventions is given in Sec.~\ref{sec:numerics}.

As $r_*\to\pm\infty$ the effective potential vanishes, $V_{\rm eff}\to0$, so Eq.~\eqref{eq:radial} reduces to $\psi''+\omega^2\psi=0$ and the solutions are plane waves. It is convenient to express the large-argument behavior with the spherical Hankel functions $h_l^{(1)}$ and $h_l^{(2)}$, which satisfy
$h_l^{(1)}(z)\sim(-i)^{\,l+1}e^{iz}/z$ and $h_l^{(2)}(z)\sim i^{\,l+1}e^{-iz}/z$ as $z\to\infty$.
Imposing purely ingoing behavior at the horizon and an incident-plus-outgoing superposition at infinity yields the boundary conditions
\begin{equation}
\psi_{\omega l}\sim
\begin{cases}
A^{\rm tr}_{\omega l}\,e^{-i\omega r_*}, & r_*\to -\infty,\\[2pt]
A^{\rm in}_{\omega l}\,e^{-i\omega r_*}+A^{\rm out}_{\omega l}\,e^{+i\omega r_*}, & r_*\to +\infty,
\end{cases}
\label{eq:bc}
\end{equation}
and flux conservation gives $|A^{\rm tr}_{\omega l}|^2+|A^{\rm out}_{\omega l}|^2=|A^{\rm in}_{\omega l}|^2$.

Following the standard partial-wave treatment of wave scattering and absorption by
spherically symmetric black holes \cite{sanchez1978elastic,sanchez1978absorption,FHMscattering,crispino2009scattering},
we impose purely ingoing behavior at the horizon and an incident-plus-outgoing decomposition at infinity,
which leads to the boundary conditions in Eq.~\eqref{eq:bc} and the associated flux relation. It is convenient to introduce the partial-wave $S$-matrix, defined by the ratio of the outgoing and ingoing amplitudes
\begin{equation}
S_l(\omega) = (-1)^{l+1}\frac{A^{\rm out}_{\omega l}}{A^{\rm in}_{\omega l}}.
\label{eq:Sdef}
\end{equation}

The absorption probability of the $l$-th mode is given by
\begin{equation}
\Gamma_l(\omega)=1-|S_l|^2=\frac{|A^{\rm tr}_{\omega l}|^2}{|A^{\rm in}_{\omega l}|^2},
\label{eq:Gamma}
\end{equation}
which characterizes the transmission of the scalar wave through the effective potential barrier.

\subsection{Absorption cross sections}
The total absorption cross section is the sum over partial contributions
\begin{equation}
    \sigma_{\rm abs}=\sum_{l=0}^\infty\sigma_{\rm abs}^{(l)},
\end{equation}
with
\begin{equation}
\sigma_{\rm abs}^{(l)}=\frac{\pi}{\omega^2} (2 l+1)\Gamma_l=\frac{\pi}{\omega^2} (2 l+1)\left(1-\left|\frac{A_{\omega l}^{\rm out}}{A_{\omega l}^{\rm in}}\right|^2\right).
\end{equation}

In the low-frequency limit only the s-wave survives, then the cross section approaches the area of BH horizon, this universality first established in Ref. \cite{das1997universality} and extended to stationary geometries in Ref. \cite{higuchi2001low}.
But in the high-frequency regime, capture dominates and the cross section rapidly oscillates around the geometric cross section $\sigma_{\rm geo}$, as we have discussed in Sec. \ref{subsec:geo}. Here we give the expression of absorption cross section in two asymptotic limits
\begin{equation}
    \sigma_{\rm abs}^{\rm lf}\approx A_H,\quad \sigma_{\rm abs}^{\rm hf}\approx \pi b_c^2-8\pi^2 \frac{\Lambda_c}{\Omega_c^3}  e^{-\pi \Lambda_c/\Omega_c}\operatorname{sinc} (2\pi b_c \omega).\label{eq:asym}
\end{equation}

\subsection{Scattering cross section}

The scattering amplitude is given by
\begin{equation}
    g(\theta)=\frac{1}{2 i \omega} \sum_{l=0}^{\infty}(2 l+1)\left[S_l(\omega)-1\right] P_l(\cos \theta)
    \label{eq:gth}
\end{equation}
and the differential scattering cross section is
\begin{equation}
    \frac{d\sigma_{\rm sc}}{d\Omega}=\bigl|g(\theta)\bigr|^2,
    \label{eq:dscs}
\end{equation}
In this description the complex quantity $S_l(\omega)$ encodes the same physical information as the
classical deflection function discussed in Sec.~\ref{sec:classical}.

Our numerical evaluation of $S_l(\omega)$ and the resulting $\sigma_{\rm abs}(\omega)$ and $d\sigma_{\rm sc}/d\Omega$, together with comparisons to the low/high-frequency limits and the classical benchmarks of Sec.~\ref{sec:classical}, are presented in the next section.

\section{Numerical Results}
\label{sec:numerics}
In this section, we present a series of numerical results concerning the absorption and scattering cross sections of massless scalar field in the background of Frolov spacetimes with variable parameters.

\subsection{Numerical setup}

Given the complexity of Frolov black holes, directly solving the horizon by setting the metric function \eqref{eq:metric} equal to zero would introduce additional computational complexity and numerical uncertainty. For this reason, unlike previous studies on absorption and scattering of BHs, we adopt the normalization by setting $r_h=1$ instead of working with the commonly used $M=1$. It is convenient to introduce the normalized parameter
\begin{equation}
	q \equiv \frac{Q}{r_h},\qquad
\ell \equiv \frac{\alpha}{r_h},\qquad
m \equiv \frac{M}{r_h},
\end{equation}
in place of the original $Q$, $\alpha$ and $M$.

\begin{figure}[H] 
    \centering 
    \includegraphics[width=0.8\textwidth]{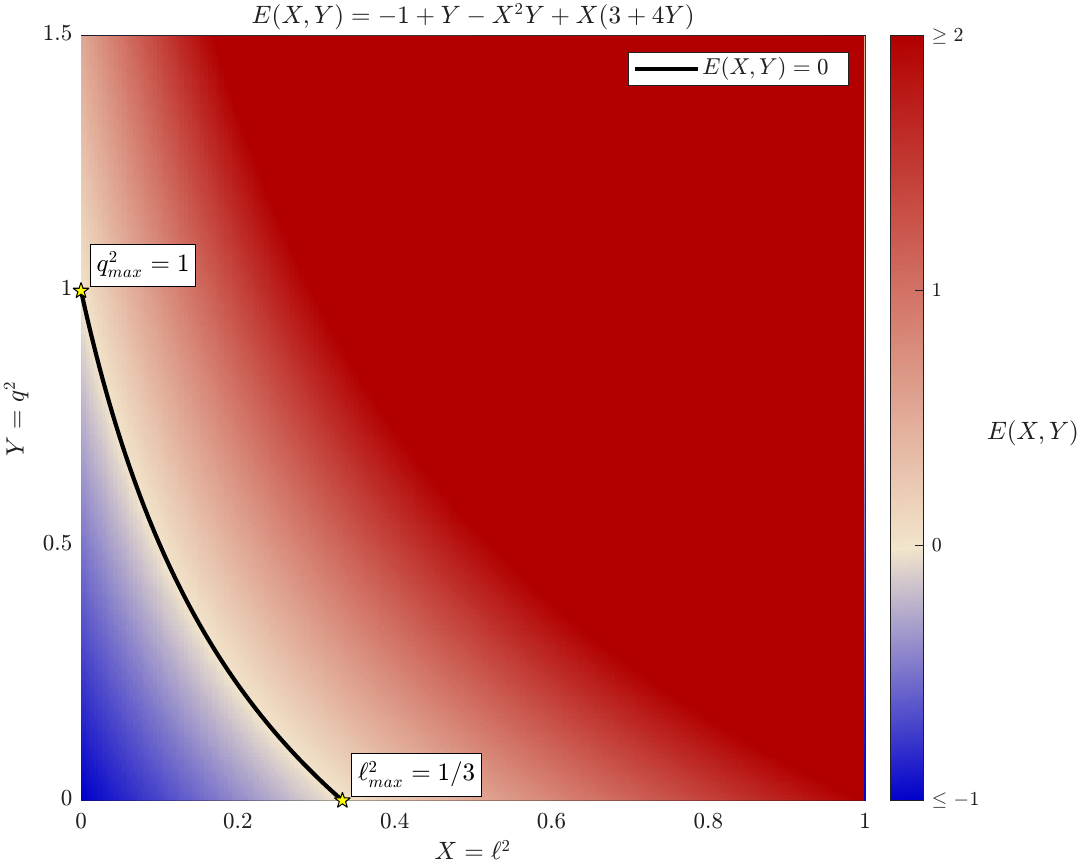}
    \caption{
Parameter space of Frolov BHs in the $(\ell^{2},q^{2})$ plane.
The blue region corresponds to configurations with an event horizon, and the solid black curve marks the extremal limit.
}
    \label{fig:parameter} 
\end{figure}

This choice imposes nontrivial constraints on the admissible values of the black hole parameters. The first constraint requires that the charge $q$ satisfy the following inequality with the Hubble length $\ell$:
\begin{equation}
    q^2-\ell^4 q^2+4\ell^2 q^2+3\ell^2\leq 1.
\end{equation}
In Fig.~\ref{fig:parameter}, we show the parameter space in the $(\ell^{2},q^{2})$ plane.  The background color encodes the function $E(\ell^{2},q^{2})=-1+q^{2}-\ell^{4}q^{2}+\ell^{2}(3+4q^{2})$, with the blue region $E\le 0$ corresponding to configurations that admit an event horizon, while the complementary region $E\ge0$ represents horizonless geometries. The solid black curve marks the locus $E=0$, where the two horizons merge and the Frolov BH becomes extremal; its intercepts at $(\ell^{2}=0,q^{2}=1)$ and $(\ell^{2}=1/3,q^{2}=0)$ reproduce, respectively, the extremal Reissner--Nordstr\"om and extremal Hayward limits. It is worth noting that, since we employ the horizon radius as our fundamental length scale, the limiting value of the regularizing parameter in the uncharged case is now $\ell=\sqrt{1/3}$, rather than the more familiar $\sqrt{16/27}$ quoted in the $M=1$ convention.

\begin{figure}[H] 
    \centering 
    \includegraphics[width=0.8\textwidth]{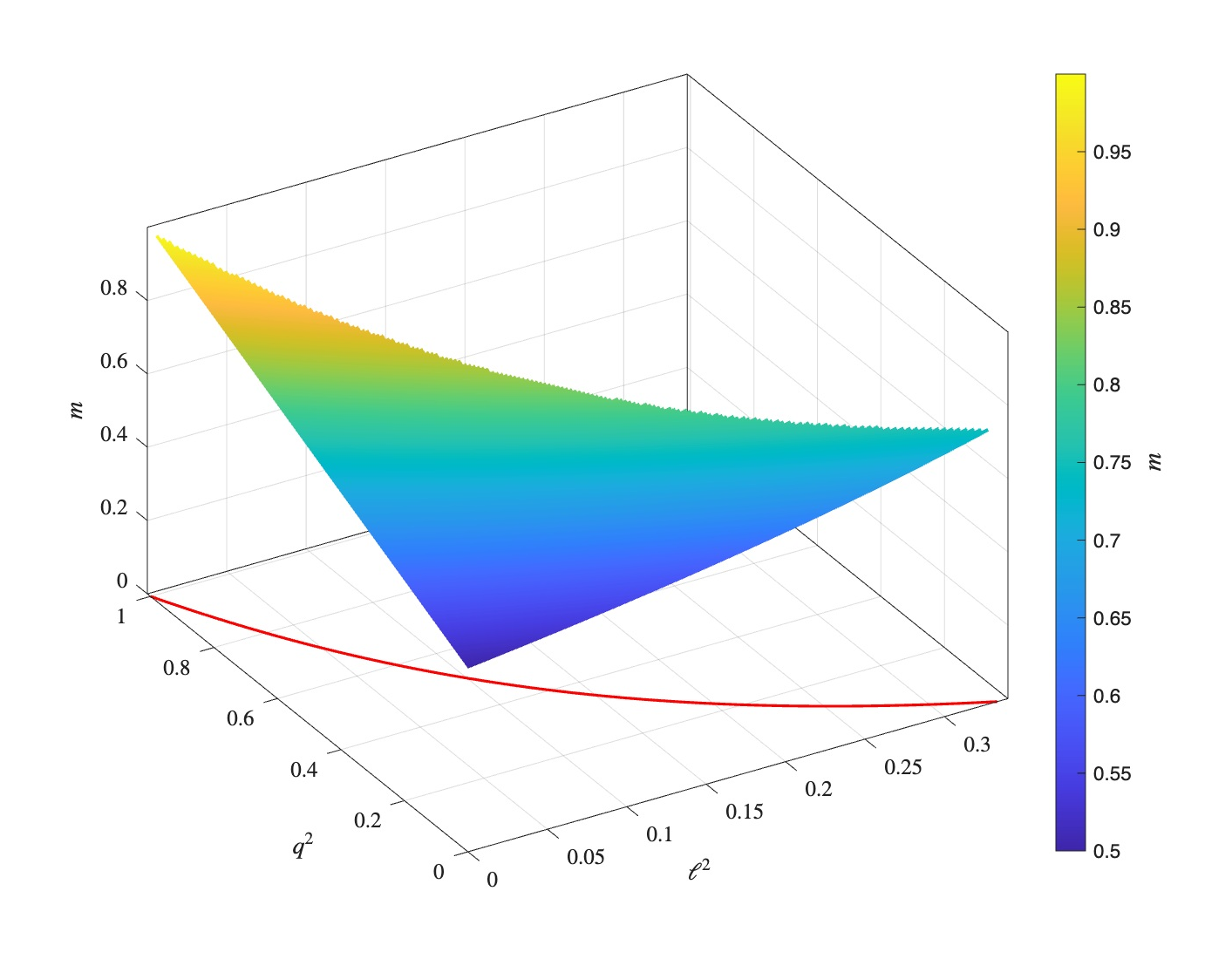}
    \caption{
Normalized mass $m$ of Frolov BHs for a fixed horizon radius $r_{h}=1$, restricted to the region admitting an event horizon.
}
    \label{fig:3dmass} 
\end{figure}

Within this domain, the condition $f(r_h=1)=0$ fixes the dimensionless mass as a function of the remaining parameters
\begin{equation}
	m(q,\ell)=\frac{1+(1+\ell^2)q^2}{2(1-\ell^2)}.
\end{equation}
Figure~\ref{fig:3dmass} displays the behaviour of $m$ as a function of $q^2$ and $\ell^2$ within the region of parameter space that admits an event horizon, which we discussed in the last paragraph. One can see that $m$ increases monotonically with both $q$ and $\ell$, so different choices of $(q,\ell)$  along the allowed domain correspond to Frolov BHs of different mass, even though the horizon radius is kept fixed. As a consequence, varying $(q,\ell)$ at fixed $r_h=1$ is not directly equivalent to varying the charge $Q$ or the Hubble length $\alpha$ at fixed mass in the usual $M=1$ normalization. This observation will be important when interpreting, in the next subsections, the dependence of the absorption and scattering cross sections on $(q,\ell)$ and when comparing our results with previous absorption scattering research in references.

Finally we numerically solve Eq. \eqref{eq:radial} from the point very close to the event horizon to a region far from the BHs, then match the numerical solutions with the boundary condition given in Eq. \eqref{eq:bc} to compute the coefficients. In calculations of absorption and scattering cross sections, we perform sums on angular momentum of the scalar wave. For the absorption case, we set $l=10$, and consider $l=50$ for the scattering case. To optimize the convergence of the differential scattering cross section at small scattering angles, we employed the well-known YRW method\cite{yennie1954phase,dolan2006fermion}. Related convergence issues of the partial-wave expansion near the optical axis, and a complementary resolution based on finite-radius wavefunctions (avoiding the problematic asymptotic expansion),
were analyzed in a rigorous manner in Ref.~\cite{li2025rigorous}.

\subsection{Scalar absorption}
\label{sec:absorption}
\begin{figure}[htb]
  \centering
  \includegraphics[width=0.48\textwidth]{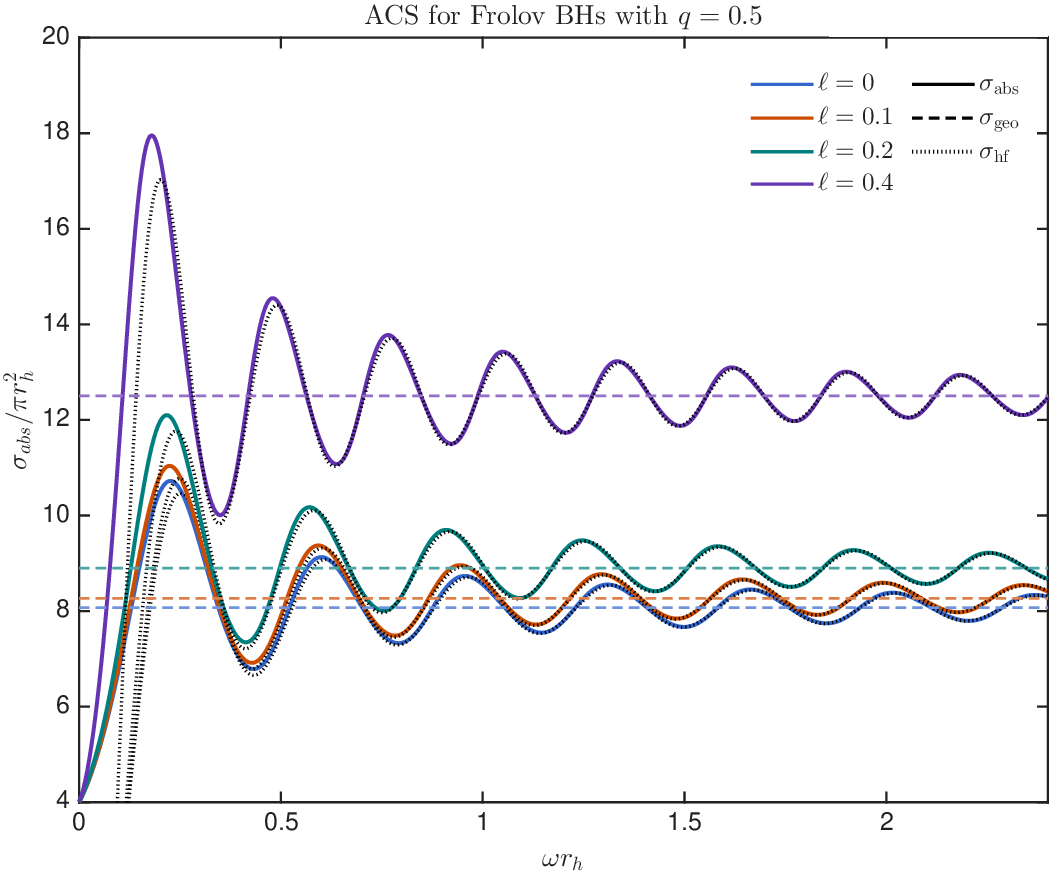}
  \includegraphics[width=0.48\textwidth]{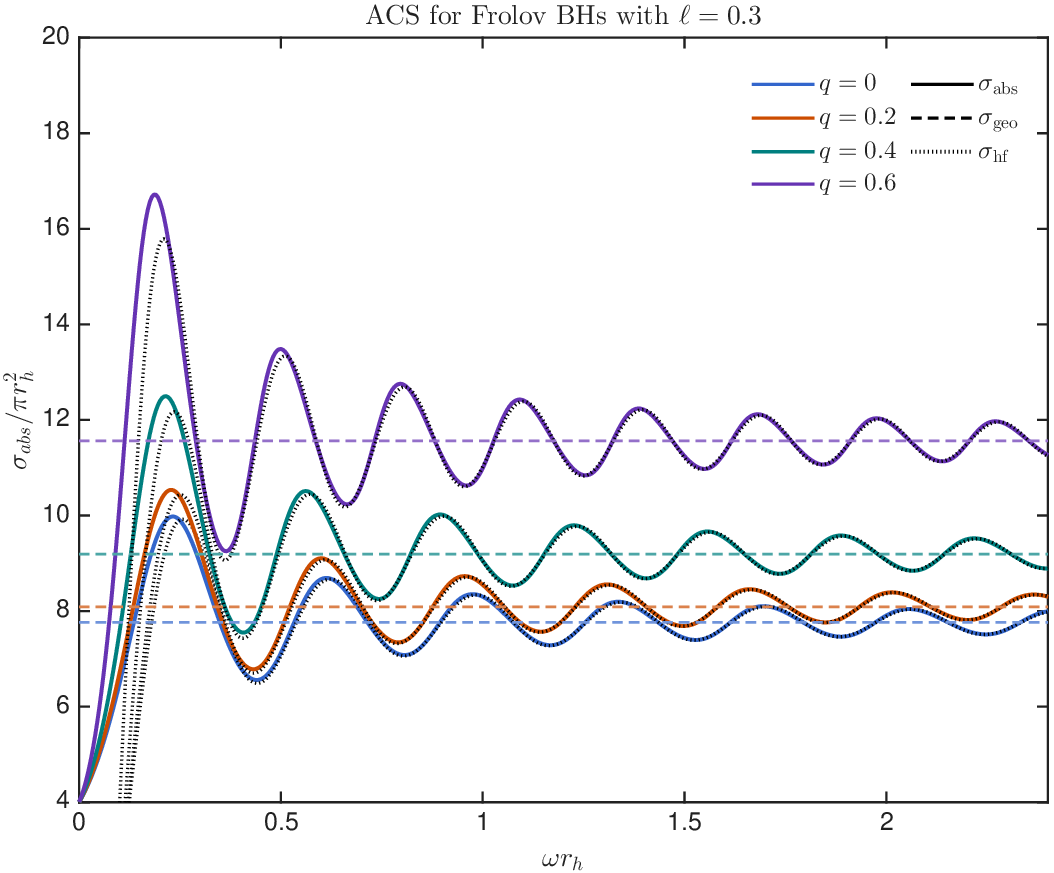}
\caption{Total absorption cross section $\sigma_{\rm abs}$ of massless scalar waves by Frolov black holes, normalized by $\pi r_h^2$, as a function of $\omega r_h$. In the left panel we fix $q=0.5$ and vary $\ell$, while in the right panel we fix $\ell=0.3$ and vary $q$. Solid curves show the numerical partial-wave result. The horizontal dashed lines indicate the geometric-optics cross section $\sigma_{\rm geo}=\pi b_c^2$ determined by the critical null geodesic. The dotted curves correspond to the high-frequency approximation $\sigma_{\rm hf}$ in Eq.~\eqref{eq:asym}, which reproduces the high-frequency oscillations of $\sigma_{\rm abs}$ around $\sigma_{\rm geo}$ with the universal Regge-pole/sinc structure controlled by photon-sphere parameters (e.g. $\Omega_c$ and the Lyapunov exponent) \cite{decanini2011universality}. In the low-frequency limit, $\sigma_{\rm abs}\to A_h=4\pi r_h^2$ as in Ref. \cite{das1997universality,higuchi2001low,sanchez1978absorption}.}
  \label{fig:totalACS}
\end{figure}

In Fig.~\ref{fig:totalACS}, we present the total absorption cross section $\sigma_{\rm abs}/(\pi r_h^2)$ of massless scalar waves for representative values of $(q,\ell)$. For reference we also display the geometric-optics limit $\sigma_{\rm geo}=\pi b_c^2$ and the high-frequency approximation $\sigma_{\rm hf}$ in Eq.~\eqref{eq:asym}.

We observe that the numerical results exhibit the standard low- and high-frequency behaviors as expected from the general theory of black-hole absorption. We also note that the parameter dependence of $\sigma_{\rm abs}/(\pi r_h^2)$ under the $r_h=1$ normalization  typically increases when either $q$ or $\ell$ is increased within the allowed domain. This trend should be interpreted with some care when compared with the common $M=1$ normalization used for RN/Hayward black holes, for which the total ACS often decreases as the charge or Hubble length grows \cite{crispino2009scattering,de2023geodesic}. In our convention the dimensionless mass $m(q,\ell)$ varies across parameter space (see Fig.~\ref{fig:3dmass}), so part of the apparent increase of $\sigma_{\rm abs}/(\pi r_h^2)$ reflects the change of the overall mass scale, in addition to the genuine variation of the photon-sphere data entering $b_c/r_h$. 

\begin{figure}[htb]
  \centering
  \includegraphics[width=0.48\textwidth]{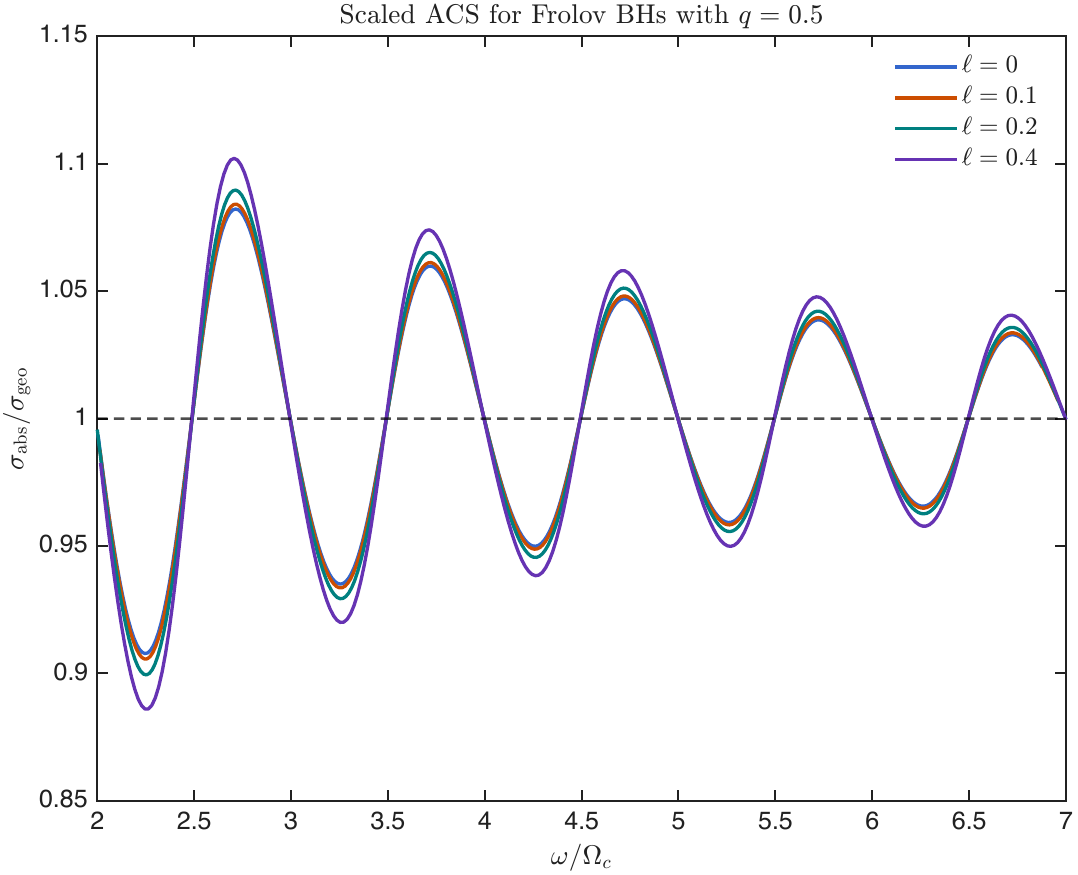}
  \includegraphics[width=0.48\textwidth]{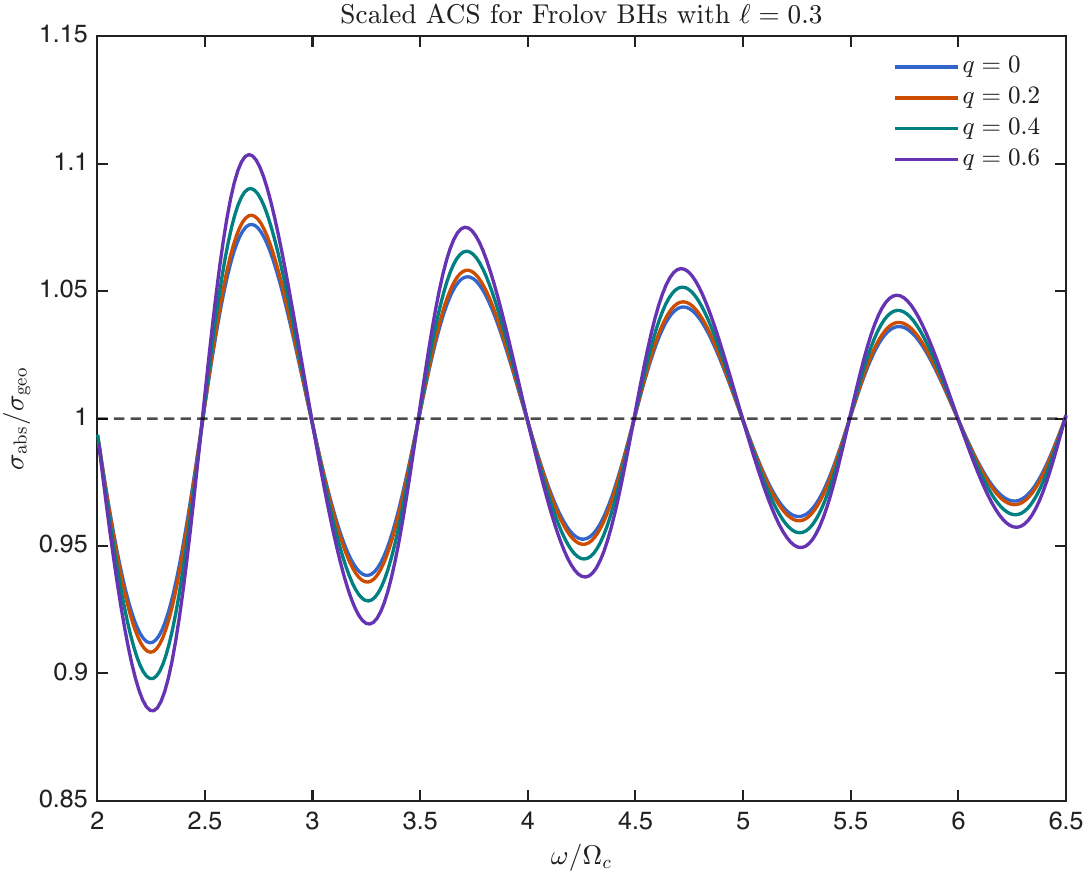}
  \caption{Scaled total absorption cross section shown as
  $\sigma_{\rm abs}/\sigma_{\rm geo}$ versus $\omega/\Omega_c$ (equivalently $\omega b_c$ for static spherical spacetimes).
  The left panel fixes $q=0.5$ and varies $\ell$, while the right panel fixes $\ell=0.3$ and varies $q$,
  matching the parameter slices of Fig.~\ref{fig:totalACS}.
  The collapse of the high-frequency oscillations in these rescaled variables highlights the photon-sphere dominance
  of the absorption fine structure (cf. the universal Regge-pole/sinc description) \cite{decanini2011universality}.}
  \label{fig:scaledACS}
\end{figure}

To disentangle the nontrivial wave effects from the overall geometric scaling induced by varying $(q,\ell)$ at fixed $r_h=1$, it is useful to re-express the absorption spectrum in terms of photon-sphere scales. We therefore consider the rescaled quantities $\hat{\sigma}\equiv \sigma_{\rm abs}/\sigma_{\rm geo}$ and $x\equiv \omega/\Omega_c$. For a static spherically symmetric geometry one has $\Omega_c=\sqrt{f(r_c)}/r_c=1/b_c$, so $x=\omega b_c$. Since $\hat{\sigma}$ and $x$ are dimensionless and invariant under an overall length rescaling, this representation does not depend on whether one adopts $r_h$ or $M$ as the fundamental unit.
As shown in Fig.~\ref{fig:scaledACS}, the high-frequency oscillations for different $(q,\ell)$ largely collapse in these variables, supporting the interpretation that the dominant high-frequency structure is controlled by photon-sphere data ($b_c$, $\Omega_c$ and the Lyapunov exponent entering the Regge-pole/sinc formula), while residual differences at intermediate frequencies reflect subleading (beyond-eikonal) effects.

\begin{figure}[htb]
  \centering
  \includegraphics[width=0.48\textwidth]{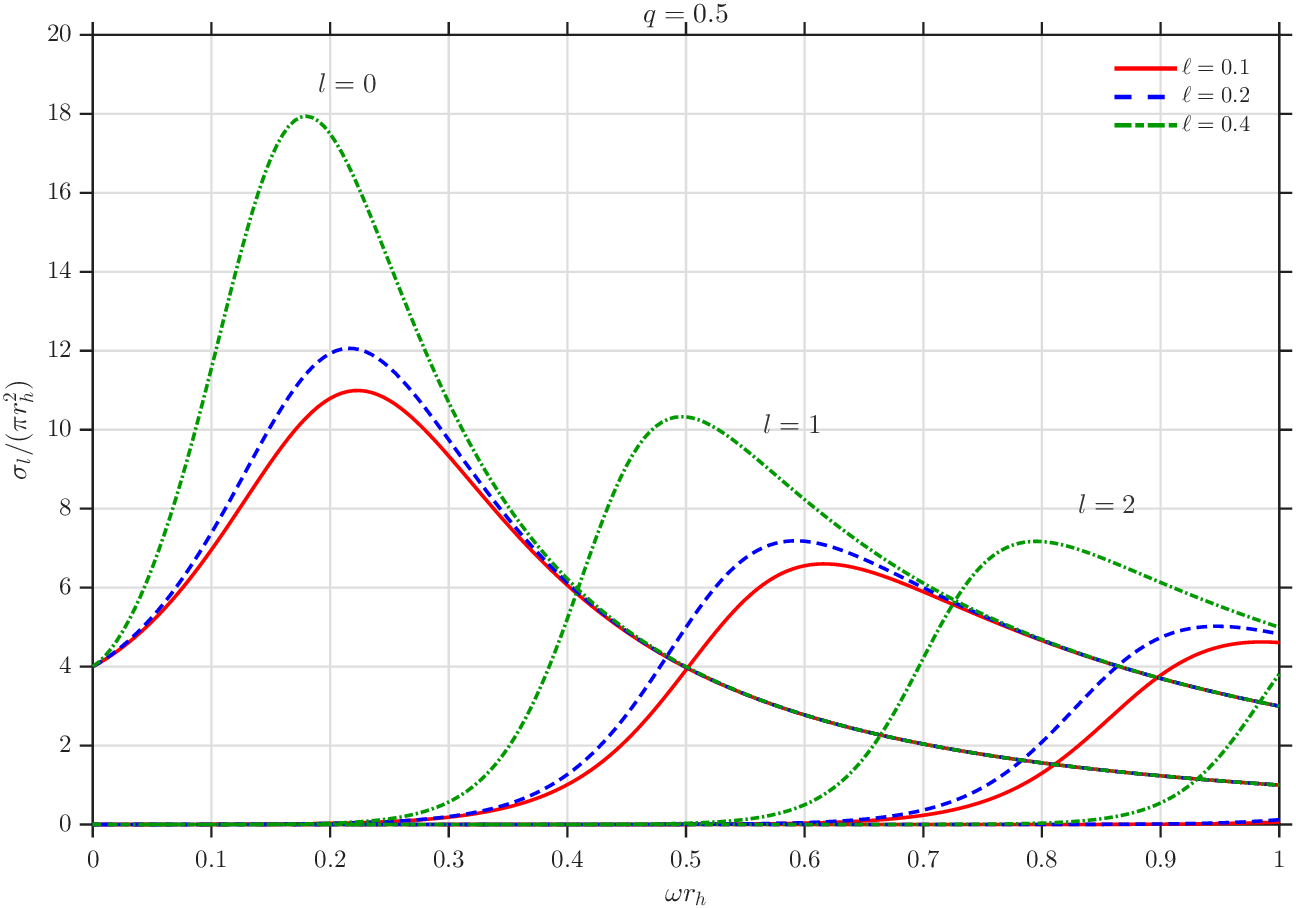}%
  \includegraphics[width=0.48\textwidth]{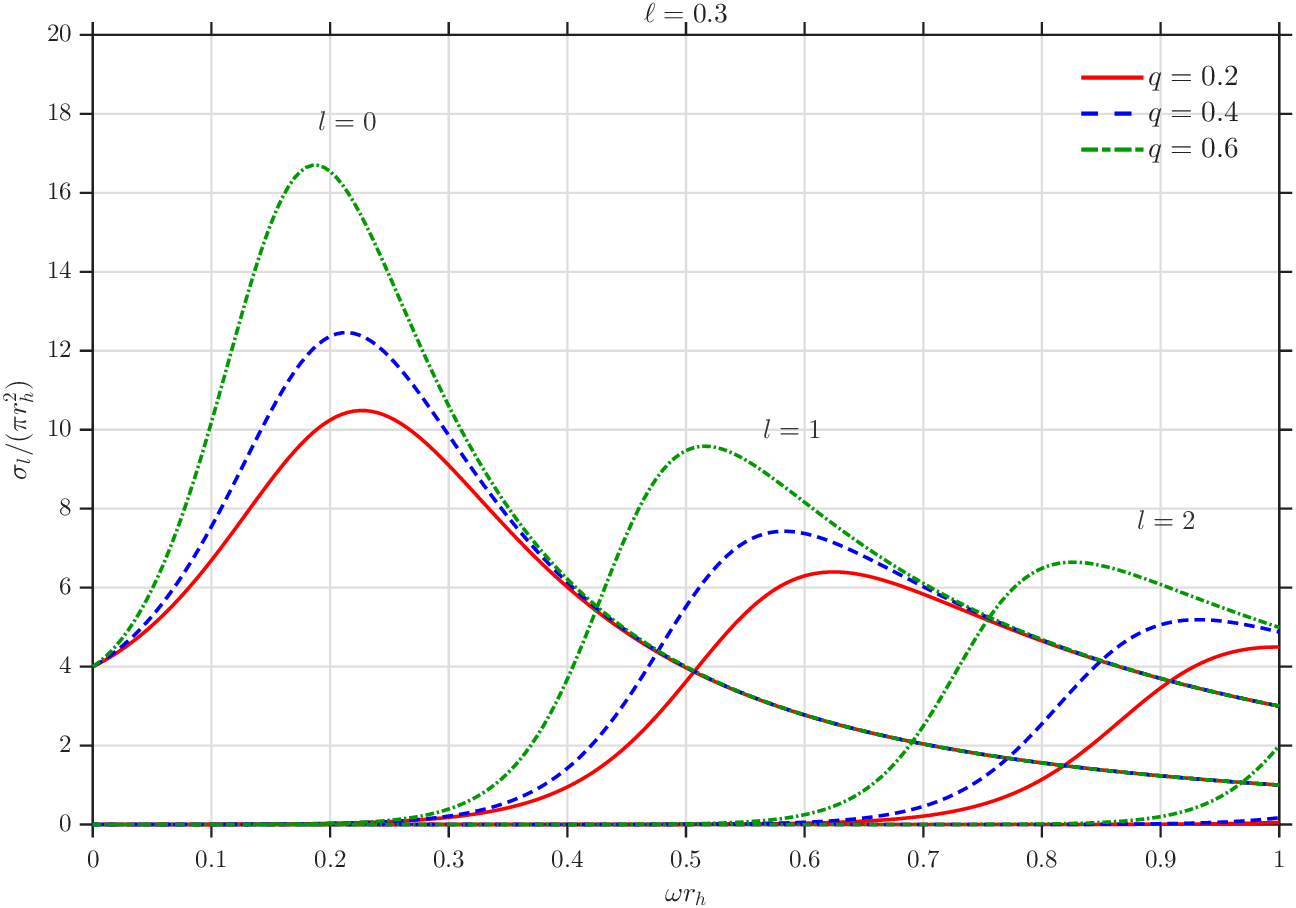}
  \caption{
Partial absorption cross sections $\sigma_l$ for the lowest modes $l=0,1,2$,
normalized by $\pi r_h^2$, as functions of $\omega r_h$. In the low-frequency limit, the absorption is dominated by the $s$-wave and $\sigma_{\rm abs}\to A_h$, whereas higher modes are suppressed by the centrifugal barrier
($\sigma_{l>0}\to0$ as $\omega\to0$) \cite{sanchez1978absorption,higuchi2001low}.
As $\omega$ increases, each partial wave is activated when the wave can penetrate the corresponding
effective-potential barrier, leading to the characteristic rise and peak structure familiar from
Schwarzschild/RN and other (regular) black holes \cite{FHMscattering,crispino2009scattering}.
  }
  \label{fig:PACS}
\end{figure}

As a complementary view, Fig.~\ref{fig:PACS} shows the partial absorption cross sections $\sigma_l$ for $l=0,1,2$ for the same parameter choices. The total ACS trend is largely controlled by the lowest modes: at low frequencies the $s$-wave dominates, while higher-$l$ channels are progressively activated only when the wave can efficiently penetrate the corresponding effective-potential barrier.

\subsection{Scattering cross section}
In this subsection we discuss the scattering of massless scalar waves by Frolov BHs. The differential scattering cross section is computed from the partial-wave series in Eq. \eqref{eq:dscs}, with phase shifts obtained from the numerical integration of the radial equation \eqref{eq:radial}.

First we compare our numerical results for the differential scattering cross sections (DCS) of massless scalar waves in the Frolov spacetime with approximations. It is clearly obtained that the DCS oscillating around the classical DCS, which we have derived in Eq. \eqref{eq:classical dcs}. The pattern agrees well with the classical analytical results in small scattering angle, and is well described by glory approximation(Eq. \eqref{eq:glory}) in the region $\theta \approx \pi$.
\begin{figure}[htb]
  \centering
  \includegraphics[width=0.8\textwidth]{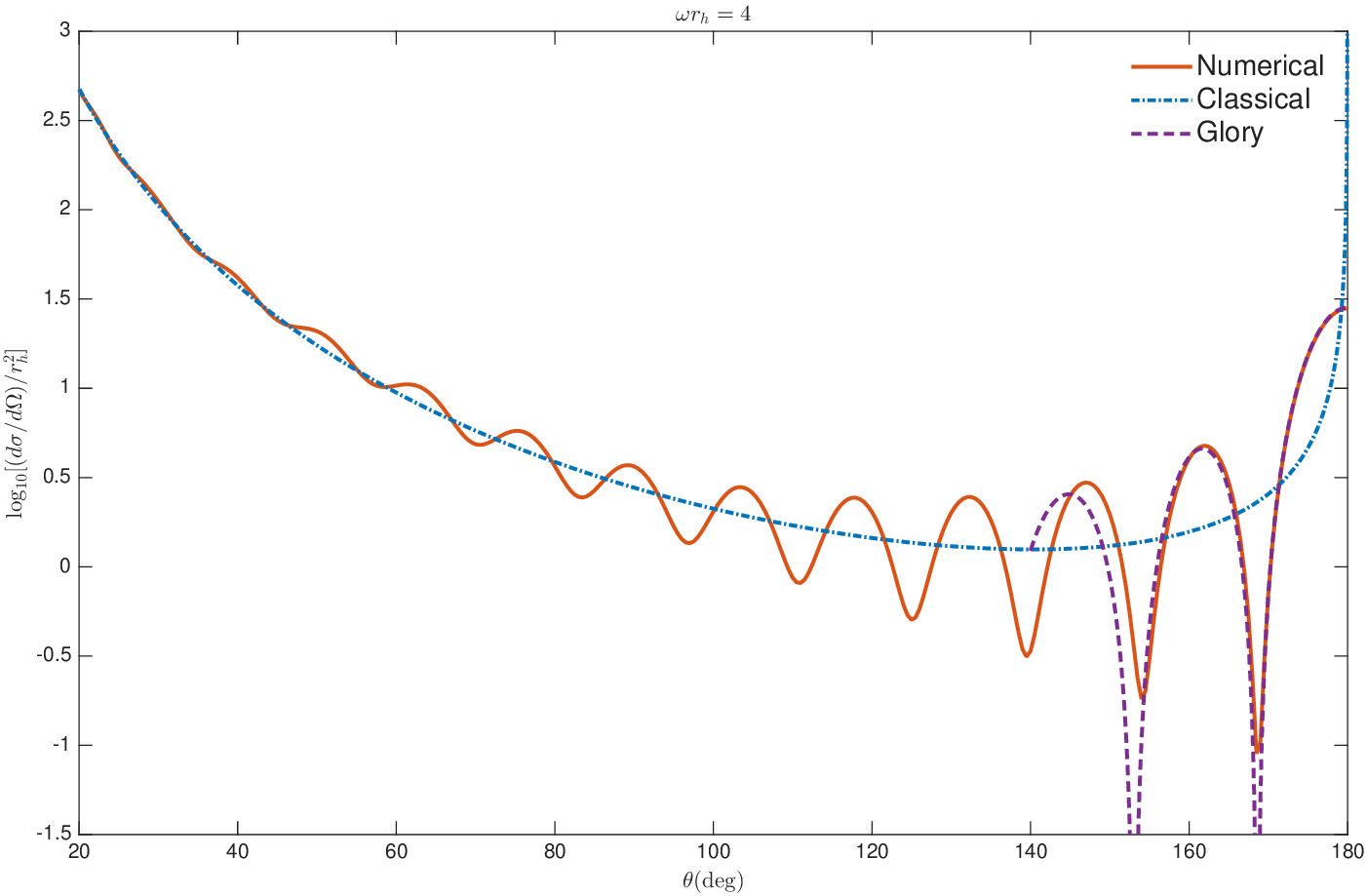}%
  \caption{
   Comparison between the numerical and the approximate analytical results for the DCS of massless scalar fields in the Frolov spacetime. Here we consider $q=0.3$ and $\ell=0.3$ as parameters for frequency $\omega r_h=4$.
  }
  \label{fig:NCGdsc}
\end{figure}

\begin{figure}[htb]
  \centering
  \includegraphics[width=0.48\textwidth]{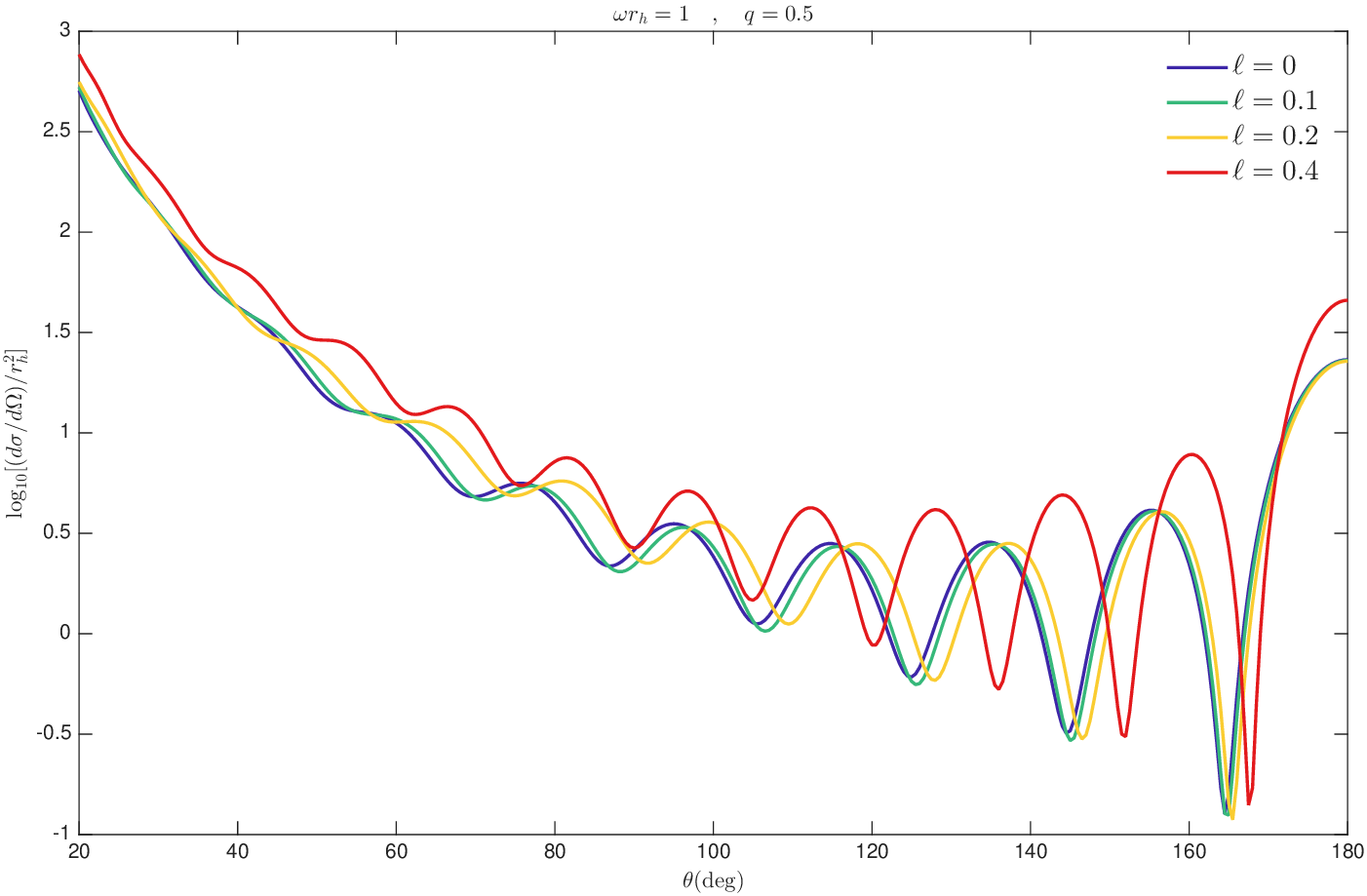}%
  \includegraphics[width=0.48\textwidth]{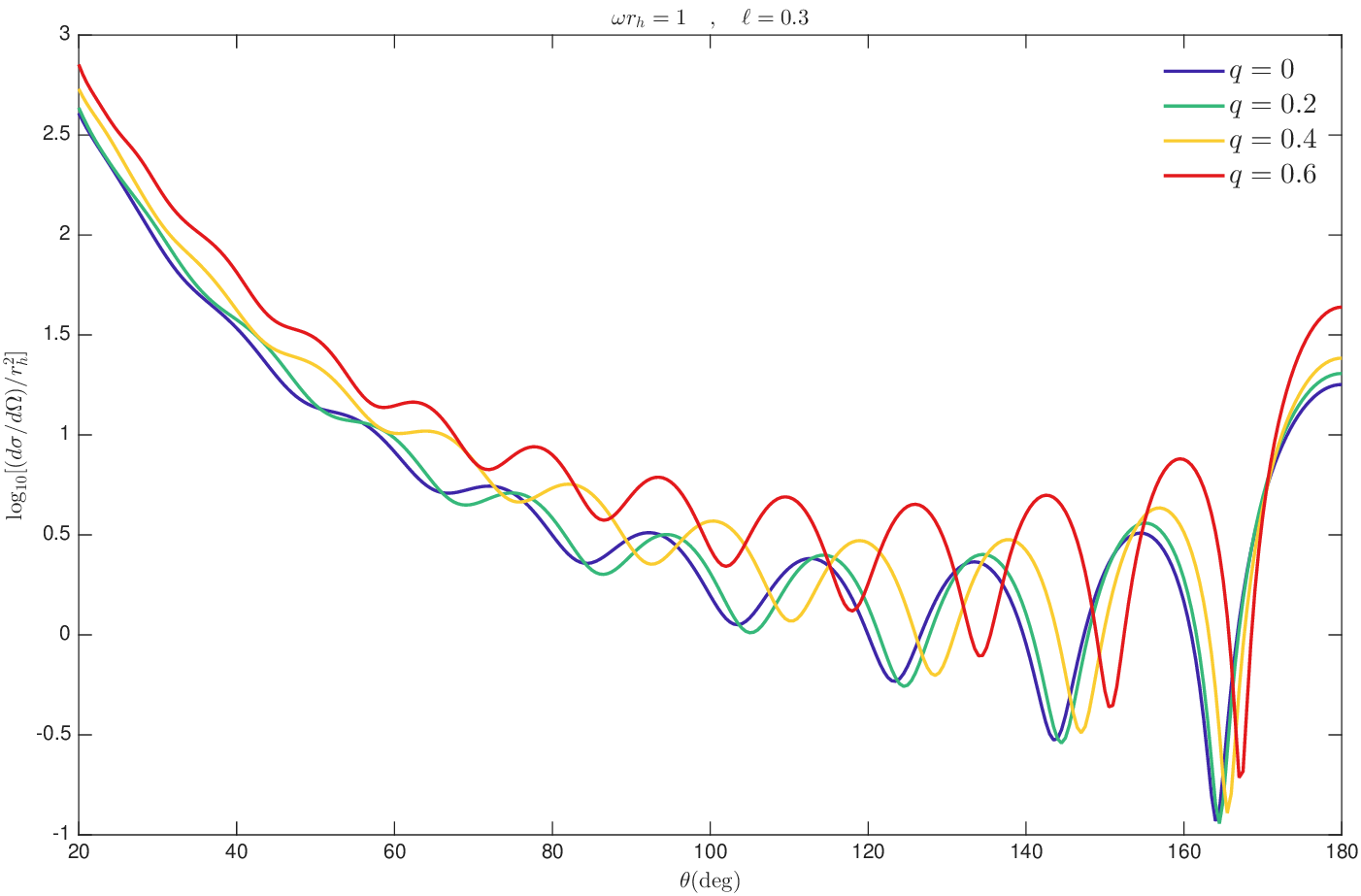}
  \caption{
    Differential scattering cross section for Frolov spacetime for different values of $q$ and $\ell$, at a representative frequency $\omega r_h =3$. 
  }
  \label{fig:dscs}
\end{figure}

Then Fig.~\ref{fig:dscs} displays the differential scattering cross sections (DCS) for massless scalar waves with frequency  $\omega r_h =3$ in the background of Frolov spacetime, which show the dependency of DCS on the parameters. The left panel shows the effect of varying $\ell$ at fixed charge, while the right panel shows the effect of varying $q$ at fixed $\ell$, where the values chosen are consistent with we used in the absorption work. One can see that the small-angle cross section is much more sensitive to changes in $q$ than to changes in $\ell$. The behavior is in agreement with the formula in Eq. \eqref{eq:classical dcs}, where the charge enters the subdominating $\Theta^{-3}$ term, whereas the dependence on $\ell$ appears only through higher-orders corrections that is neglected in the approximation. However at large angles, the effect of $\ell$ becomes comparable to that of $q$, which they have a significant effect upon DCS. We also found that the interference fringe width decreases with the increase of two parameters $q$ and $\ell$ for given frequency $\omega r_h$. This behavior is opposite to what is typically found for Reissner--Nordstr\"om and Hayward black holes when the mass is kept fixed\cite{crispino2009scattering,de2023geodesic}. This difference arises from our choice of units, where the horizon radius rather than mass is used as the fundamental scale.

\subsection{Similar features for different Frolov BHs}

The intriguing work of de Paula et al.~\cite{de2023geodesic} investigated under which circumstances the absorption and scattering spectra of Hayward and Reissner--Nordstr\"om (RN) black holes can become nearly indistinguishable. In the absorption problem, they matched the critical impact parameters by tuning the model parameters, $b_c^{\rm H}(\alpha)=b_c^{\rm RN}(Q)$, and found that the corresponding total absorption cross sections (ACSs) are very similar over a broad frequency range. In the scattering problem, they instead matched the impact parameter of the backscattered null geodesics, $b_g^{\rm H}(\alpha)=b_g^{\rm RN}(Q)$, and observed that the differential scattering cross sections (DCSs) exhibit nearly identical interference patterns.

Motivated by this observation, we ask whether a similar correspondence persists when comparing RN, Hayward, and Frolov black holes. Since the Frolov spacetime reduces to RN and Hayward geometries in appropriate limits, curves of constant impact parameters $b(\ell,q)$ in the $(\ell^2,q^2)$ plane (cf. Fig.~\ref{fig:parameter}) naturally interpolate between these limiting cases. Along a given iso-$b$ curve there therefore exists a continuous family of Frolov black holes sharing the same characteristic impact parameter. 

\begin{figure}[htb]
  \centering
  \includegraphics[width=0.48\textwidth]{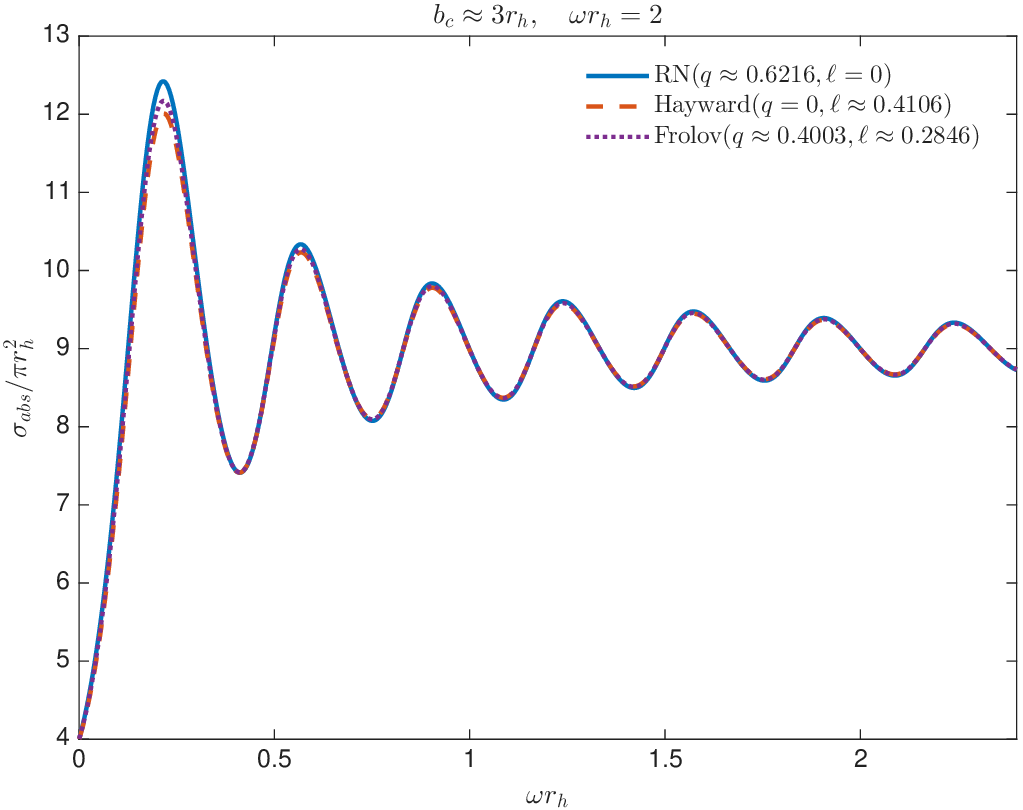}\hfill
  \includegraphics[width=0.48\textwidth]{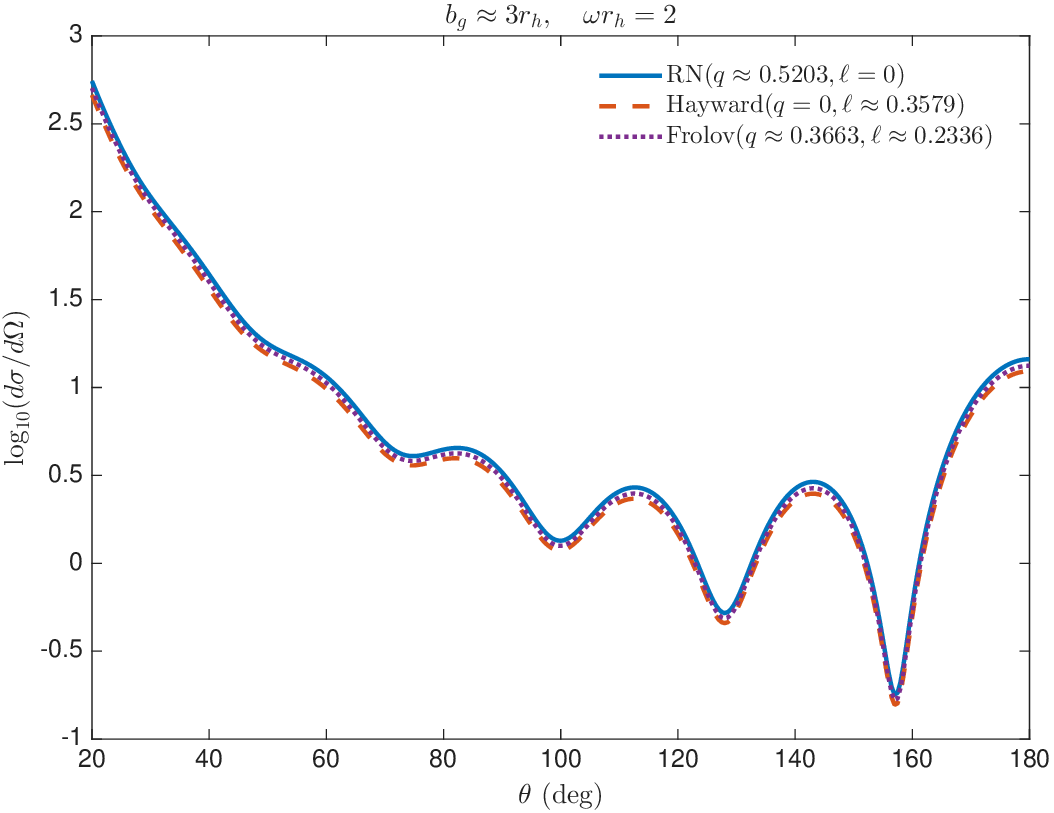}
  \caption{ACSs and DCSs of Reissner--Nordstr\"om, Hayward and Frolov black holes with matched impact parameters.
  The left panel compares ACSs for configurations with $b_c\simeq 3\,r_h$, while the right panel compares DCSs
  for configurations with $b_g\simeq 3\,r_h$.}
  \label{fig:iso_b_comparison}
\end{figure}

To assess the extent to which the wave observables are controlled by these impact parameters,we focus on the representative lines $b_c\simeq 3\,r_h$ and $b_g\simeq 3\,r_h$ and select three configurations on each line:
an RN black hole, a Hayward black hole, and an intermediate Frolov case. Figure~\ref{fig:iso_b_comparison} compares the corresponding ACSs (left panel, matched $b_c$) and DCSs (right panel, matched $b_g$).

We observe that, under this matching strategy, the ACSs and DCSs of the three geometries nearly coincide over the displayed frequency/angle ranges. While the agreement in the asymptotic limits is expected (the low-frequency behavior is governed by $r_h$, whereas the high-frequency structure is controlled by the photon-sphere data encoded in $b_c$ or $b_g$), the close overlap in the intermediate regime is nontrivial. This suggests that the relevant effective potentials possess very similar barrier shapes in the vicinity of the photon sphere, so that the dominant features of massless-scalar absorption and scattering are largely controlled by the unstable null orbit, with the detailed regular-core structure contributing only subleading corrections in these configurations. Accordingly, within the present probe (massless scalar) and the parameter-matched setup, distinguishing the three spacetimes using these wave observables alone can be challenging.

\section{Conclusion}
\label{sec:conclusion}
The absorption and scattering of fields by black holes serve as a powerful probe for exploring the fundamental nature of spacetime geometry and horizon structure. Regular black holes, which smooth out central singularities, have been shown to leave distinct signatures in the scalar absorption cross sections and differential scattering cross sections across several specific models\cite{macedo2014absorption,paula2020electrically}. However, a systematic investigation into how the two key parameters of the Frolov spacetime jointly influence these observable quantities over a broad frequency range has remained lacking prior to this work.

In this work, we first conducted a comprehensive null geodesic analysis to confirm the existence of the event horizon in Frolov black holes, pinpoint the location of the photon sphere, and calculate the critical impact parameters that govern the capture process and glory scattering. On the wave physics front, we employed the partial-wave method to investigate minimally coupled massless scalar fields. By solving the radial equation with appropriately imposed boundary conditions, we derived the partial absorption probabilities, total and partial absorption cross sections, as well as the differential scattering cross section. Across the entire parameter space explored, the numerically computed cross sections smoothly bridge the universal low-frequency regime and the high-frequency geometric optics regime, demonstrating excellent agreement with classical approximations and glory approximations within their respective valid domains.

In the numerical analysis we adopted the normalization $r_h=1$ and mapped the allowed region in the $(\ell^2,q^2)$ plane where an event horizon exists. In this parametrization the normalized mass $m$ grows monotonically with both $q$ and $\ell$, so changing these parameters at fixed $r_h$ does not correspond to the same physical variation as in the more common $M=1$ convention. This explains why, in our results, the total absorption cross section and the height of the effective potential barrier increase as either parameter is raised, even though related studies of Reissner--Nordstr\"om and Hayward black holes at fixed mass report a decreasing trend. The partial-wave decomposition further shows that the lowest mode dominates the low-frequency absorption and drives most of the parameter dependence, while varying $(q,\ell)$ in the scattering problem mainly shifts and distorts the interference fringe pattern.

We also presented a scale-independent characterization of the absorption spectrum by rescaling the results with photon-sphere scales. Using $\hat{\sigma}=\sigma_{\rm abs}/\sigma_{\rm geo}$ and $x=\omega/\Omega_c$, we found that the high-frequency oscillations for different $(q,\ell)$ largely collapse onto a nearly universal curve, see Fig.~\ref{fig:scaledACS}. Because both $\hat{\sigma}$ and $x$ are dimensionless and invariant under an overall length rescaling, this collapse is insensitive to whether one adopts $r_h$ or $M$ as the fundamental unit. It highlights that the dominant high-frequency structure is primarily encoded in photon-sphere data, while residual differences at intermediate frequencies reflect subleading effects beyond the eikonal description.

Motivated by recent comparisons between different regular metrics at fixed impact parameters, we also examined Frolov, Reissner--Nordstr\"om and Hayward black holes chosen to share the same critical impact parameter $b_c$ for absorption and the glory parameter $b_g$ for scattering.
Along such iso--$b_c$ and iso--$b_g$ curves in parameter space, the effective potentials and the resulting ACS and DCS are found to be remarkably close over the displayed frequency and angle ranges, including the intermediate regime where no a priori guarantee of overlap exists. This provides concrete evidence that, within this class of spacetimes and for massless scalar probes, the dominant features of absorption and scattering are largely controlled by the unstable photon orbit, while the detailed regular-core structure leaves only subleading imprints on these observables. Overall, the rescaled-spectrum plot demonstrates photon-sphere control \emph{within} the Frolov family under parameter variations, whereas the iso--$b$ comparisons extend this idea \emph{across} different regular metrics through impact-parameter matching.

Several promising directions for extending the present work naturally emerge. First, investigating higher-spin fields or nonminimally coupled scalar fields in the Frolov spacetime would help assess the universality of the patterns identified herein. Second, exploring rotating generalizations of regular black holes could uncover qualitatively new signatures in absorption and scattering, as the interplay between regular cores and frame-dragging effects may introduce novel physical behaviors. Such extensions would further deepen our understanding of the interaction between waves and regular black holes, providing valuable insights for future observational tests of gravitational theories. Beyond the present static setup, it is natural to consider charged/massive scalar probes in charged backgrounds; see, e.g., Ref.~\cite{li2025absorption} and Ref.~\cite{li2025scattering}. It would also be interesting to connect black hole scattering/absorption calculations to wave-optics gravitational-wave lensing,
where spin effects can enter naturally \cite{li2025gravitational}.

\section*{ACKNOWLEDGMENTS}

This work is supported by the National Natural Science Foundation of China (NSFC) under Grant nos. 12235019, 12275106 and Shandong Provincial Natural Science Foundation under grant No. ZR2024QA032.

\bibliographystyle{unsrt}
\bibliography{fro}

@article{das1997universality,
  title={Universality of low energy absorption cross sections for black holes},
  author={Das, Sumit R and Gibbons, Gary and Mathur, Samir D},
  journal={Physical Review Letters},
  volume={78},
  number={3},
  pages={417},
  year={1997},
  publisher={APS}
}

@article{EHT2019I,
  author       = {{Event Horizon Telescope Collaboration}},
  title        = {First M87 Event Horizon Telescope Results. I. The Shadow of the Supermassive Black Hole},
  journal      = {Astrophysical Journal Letters},
  year         = {2019},
  volume       = {875},
  number       = {1},
  pages        = {L1},
  doi          = {10.3847/2041-8213/ab0ec7}
}

@article{EHT2022I,
  author       = {{Event Horizon Telescope Collaboration}},
  title        = {First Sagittarius A* Event Horizon Telescope Results. I. The Shadow of the Supermassive Black Hole in the Center of the Milky Way},
  journal      = {Astrophysical Journal Letters},
  year         = {2022},
  volume       = {930},
  number       = {2},
  pages        = {L12},
  doi          = {10.3847/2041-8213/ac6674}
}

@article{higuchi2001low,
  title={Low-frequency scalar absorption cross sections forstationary black holes},
  author={Higuchi, Atsushi},
  journal={Classical and Quantum Gravity},
  volume={18},
  number={20},
  pages={L139},
  year={2001},
  publisher={IOP Publishing}
}

@article{higuchi2002addendum,
  title={Addendum to ‘Low-frequency scalar absorption cross sections for stationary black holes’},
  author={Higuchi, Atsushi},
  journal={Classical and Quantum Gravity},
  volume={19},
  number={3},
  pages={599},
  year={2002},
  publisher={IOP Publishing}
}

@article{decanini2011universality,
  title={Universality of high-energy absorption cross sections for black holes},
  author={D{\'e}canini, Yves and Esposito-Farese, Gilles and Folacci, Antoine},
  journal={Physical Review D—Particles, Fields, Gravitation, and Cosmology},
  volume={83},
  number={4},
  pages={044032},
  year={2011},
  publisher={APS}
}

@article{sanchez1978absorption,
  title={Absorption and emission spectra of a Schwarzschild black hole},
  author={Sanchez, Norma},
  journal={Physical Review D},
  volume={18},
  number={4},
  pages={1030},
  year={1978},
  publisher={APS}
}

@article{sanchez1978elastic,
  title={Elastic scattering of waves by a black hole},
  author={S{\'a}nchez, Norma},
  journal={Physical Review D},
  volume={18},
  number={6},
  pages={1798},
  year={1978},
  publisher={APS}
}

@article{frolov2016notes,
  title={Notes on nonsingular models of black holes},
  author={Frolov, Valeri P},
  journal={Physical Review D},
  volume={94},
  number={10},
  pages={104056},
  year={2016},
  publisher={APS}
}

@book{FHMscattering,
  author       = {Futterman, J A.H. and Handler, F A and Matzner, R A},
  title        = {Scattering from black holes},
  url          = {https://www.osti.gov/biblio/5598738},
  place        = {United States},
  publisher    = {Cambridge University Press,New York, NY},
  year         = {1986},
  month        = {12}}

@article{crispino2009scattering,
  title={Scattering of massless scalar waves by Reissner-Nordstr{\"o}m black holes},
  author={Crispino, Lu{\'\i}s CB and Dolan, Sam R and Oliveira, Ednilton S},
  journal={Physical Review D—Particles, Fields, Gravitation, and Cosmology},
  volume={79},
  number={6},
  pages={064022},
  year={2009},
  publisher={APS}
}

@article{Kala:2025FrolovRBH,
  author       = {Kala, Shubham and Nandan, Hemwati and Maithani, Kush and Roy, Saswati and Abebe, Amare},
  title        = {Null geodesics, thermodynamics, weak gravitational lensing, and black hole shadow characteristics of a Frolov regular black hole with constraints from EHT observations},
  journal      = {The European Physical Journal Plus},
  year         = {2025},
  volume       = {140},
  pages        = {991},             
  doi          = {10.1140/epjp/s13360-025-06930-9},
  url          = {https://link.springer.com/article/10.1140/epjp/s13360-025-06930-9}
}

@article{de2023geodesic,
  title={Geodesic analysis, absorption and scattering in the static Hayward spacetime},
  author={de Paula, Marco AA and Leite, Luiz and Crispino, Lu{\'\i}s CB},
  journal={arXiv preprint arXiv:2311.15771},
  year={2023}
}

@book{chandrasekhar1998mathematical,
  title={The mathematical theory of black holes},
  author={Chandrasekhar, Subrahmanyan},
  volume={69},
  year={1998},
  publisher={Oxford university press}
}

@book{newton2013scattering,
  title={Scattering theory of waves and particles},
  author={Newton, Roger G},
  year={2013},
  publisher={Springer Science \& Business Media}
}

@article{matzner1985glory,
  title={Glory scattering by black holes},
  author={Matzner, Richard A and DeWitte-Morette, C{\'e}cile and Nelson, Bruce and Zhang, Tian-Rong},
  journal={Physical Review D},
  volume={31},
  number={8},
  pages={1869},
  year={1985},
  publisher={APS}
}

@article{macedo2015scattering,
  title={Scattering by regular black holes: planar massless scalar waves impinging upon a Bardeen black hole},
  author={Macedo, Caio FB and de Oliveira, Ednilton S and Crispino, Lu{\'\i}s CB},
  journal={Physical Review D},
  volume={92},
  number={2},
  pages={024012},
  year={2015},
  publisher={APS}
}

@article{de2022scattering,
  title={Scattering properties of charged black holes in nonlinear and Maxwell’s electrodynamics},
  author={de Paula, Marco AA and Leite, Luiz and Crispino, Lu{\'\i}s CB},
  journal={The European Physical Journal Plus},
  volume={137},
  number={7},
  pages={1--13},
  year={2022},
  publisher={Springer}
}

@article{murk2023regular,
  title={Regular black holes and the first law of black hole mechanics},
  author={Murk, Sebastian and Soranidis, Ioannis},
  journal={Physical Review D},
  volume={108},
  number={4},
  pages={044002},
  year={2023},
  publisher={APS}
}

@article{song2024quasinormal,
  title={Quasinormal modes and ringdown waveforms of a Frolov black hole},
  author={Song, Zhijun and Gong, Huajie and Li, Hai-Li and Fu, Guoyang and Zhu, Li-Gang and Wu, Jian-Pin},
  journal={Communications in Theoretical Physics},
  volume={76},
  number={10},
  pages={105401},
  year={2024},
  publisher={IOP Publishing}
}

@book{Wald:1984,
  author    = {Wald, Robert M.},
  title     = {General Relativity},
  publisher = {University of Chicago Press},
  address   = {Chicago},
  year      = {1984},
  isbn      = {978-0226870335}
}

@article{cardoso2009geodesic,
  title={Geodesic stability, Lyapunov exponents, and quasinormal modes},
  author={Cardoso, Vitor and Miranda, Alex S and Berti, Emanuele and Witek, Helvi and Zanchin, Vilson T},
  journal={Physical Review D—Particles, Fields, Gravitation, and Cosmology},
  volume={79},
  number={6},
  pages={064016},
  year={2009},
  publisher={APS}
}

@article{yennie1954phase,
  title={Phase-shift calculation of high-energy electron scattering},
  author={Yennie, DR and Ravenhall, D\_ G\_ and Wilson, RN},
  journal={Physical Review},
  volume={95},
  number={2},
  pages={500},
  year={1954},
  publisher={APS}
}

@article{dolan2006fermion,
  title={Fermion scattering by a Schwarzschild black hole},
  author={Dolan, Sam and Doran, Chris and Lasenby, Anthony},
  journal={Physical Review D—Particles, Fields, Gravitation, and Cosmology},
  volume={74},
  number={6},
  pages={064005},
  year={2006},
  publisher={APS}
}

@article{ansoldi2008spherical,
  title={Spherical black holes with regular center: a review of existing models including a recent realization with Gaussian sources},
  author={Ansoldi, Stefano},
  journal={arXiv preprint arXiv:0802.0330},
  year={2008}
}

@article{ayon1998regular,
  title={Regular black hole in general relativity coupled to nonlinear electrodynamics},
  author={Ayon-Beato, Eloy and Garcia, Alberto},
  journal={Physical review letters},
  volume={80},
  number={23},
  pages={5056},
  year={1998},
  publisher={APS}
}

@article{ayon2000bardeen,
  title={The Bardeen model as a nonlinear magnetic monopole},
  author={Ayon-Beato, Eloy and Garc{\i}a, Alberto},
  journal={Physics Letters B},
  volume={493},
  number={1-2},
  pages={149--152},
  year={2000},
  publisher={Elsevier}
}

@article{balart2014regular,
  title={Regular black holes with a nonlinear electrodynamics source},
  author={Balart, Leonardo and Vagenas, Elias C},
  journal={Physical Review D},
  volume={90},
  number={12},
  pages={124045},
  year={2014},
  publisher={APS}
}

@inproceedings{bardeen1968non,
  title={Non-singular general relativistic gravitational collapse},
  author={Bardeen, James},
  booktitle={Proceedings of the 5th International Conference on Gravitation and the Theory of Relativity},
  pages={87},
  year={1968}
}

@article{hayward2006formation,
  title={Formation and evaporation of nonsingular black holes},
  author={Hayward, Sean A},
  journal={Physical review letters},
  volume={96},
  number={3},
  pages={031103},
  year={2006},
  publisher={APS}
}

@article{macedo2014absorption,
  title={Absorption of planar massless scalar waves by Bardeen regular black holes},
  author={Macedo, Caio FB and Crispino, Lu{\'\i}s CB},
  journal={Physical Review D},
  volume={90},
  number={6},
  pages={064001},
  year={2014},
  publisher={APS}
}

@article{paula2020electrically,
  title={Electrically charged black holes in linear and nonlinear electrodynamics: Geodesic analysis and scalar absorption},
  author={Paula, Marco AA and Leite, Luiz CS and Crispino, Luis CB},
  journal={Physical Review D},
  volume={102},
  number={10},
  pages={104033},
  year={2020},
  publisher={APS}
}

@article{abbott2016observation,
  title={Observation of gravitational waves from a binary black hole merger},
  author={Abbott, Benjamin P and Abbott, Richard and Abbott, Thomas D and Abernathy, Matthew R and Acernese, Fausto and Ackley, Kendall and Adams, Carl and Adams, Thomas and Addesso, Paolo and Adhikari, Rana X and others},
  journal={Physical review letters},
  volume={116},
  number={6},
  pages={061102},
  year={2016},
  publisher={APS}
}

@incollection{anderson2002scattering,
  title={Scattering by black holes},
  author={Anderson, Nils and Jensen, Bruce},
  booktitle={Scattering},
  pages={1607--1626},
  year={2002},
  publisher={Elsevier}
}

@book{maggiore2008gravitational1,
  title={Gravitational waves: Volume 1: Theory and experiments},
  author={Maggiore, Michele},
  volume={1},
  year={2008},
  publisher={Oxford university press}
}

@book{maggiore2008gravitational2,
  title={Gravitational Waves: Astrophysics and Cosmology},
  author={Maggiore, Michele},
  volume={2},
  year={2008},
  publisher={Oxford University Press}
}

@book{creighton2012gravitational,
  title={Gravitational-wave physics and astronomy: An introduction to theory, experiment and data analysis},
  author={Creighton, Jolien DE and Anderson, Warren G},
  year={2012},
  publisher={John Wiley \& Sons}
}

@article{sathyaprakash2009physics,
  title={Physics, astrophysics and cosmology with gravitational waves},
  author={Sathyaprakash, Bangalore Suryanarayana and Schutz, Bernard F},
  journal={Living reviews in relativity},
  volume={12},
  number={1},
  pages={2},
  year={2009},
  publisher={Springer}
}

@article{hughes2003listening,
  title={Listening to the universe with gravitational-wave astronomy},
  author={Hughes, Scott A},
  journal={Annals of Physics},
  volume={303},
  number={1},
  pages={142--178},
  year={2003},
  publisher={Elsevier}
}

@book{bambi2023regular,
  title={Regular black holes},
  author={Bambi, Cosimo},
  year={2023},
  publisher={Springer}
}

@article{bronnikov2007regular,
  title={Regular black holes and black universes},
  author={Bronnikov, KA and Dehnen, H and Melnikov, VN},
  journal={General Relativity and Gravitation},
  volume={39},
  number={7},
  pages={973--987},
  year={2007},
  publisher={Springer}
}

@article{lan2023regular,
  title={Regular black holes: a short topic review},
  author={Lan, Chen and Yang, Hao and Guo, Yang and Miao, Yan-Gang},
  journal={International Journal of Theoretical Physics},
  volume={62},
  number={9},
  pages={202},
  year={2023},
  publisher={Springer}
}

@article{li2025scattering,
  title={Scattering of charged massive scalar waves by Kerr-Newman black holes},
  author={Li, Qian and Wang, Qianchuan and Jia, Junji},
  journal={arXiv preprint arXiv:2511.21318},
  year={2025}
}

@article{huang2025dynamics,
  title={Dynamics of photons and shadows for black holes haired with parity-odd fields},
  author={Huang, Yang and Liu, Dao-Jun and Zhang, Hongsheng},
  journal={Journal of High Energy Physics},
  volume={2025},
  number={11},
  pages={1--18},
  year={2025},
  publisher={Springer}
}

@article{huang2025lensing,
  title={Lensing and light rings of parity-odd rotating boson stars},
  author={Huang, Yang and Liu, Dao-Jun and Zhang, Hongsheng},
  journal={Science China Physics, Mechanics \& Astronomy},
  volume={68},
  number={8},
  pages={280411},
  year={2025},
  publisher={Springer}
}

@article{zhang2021spherical,
  title={Spherical gravitational waves and quasi-spherical waves scattered from black string in massive gravity},
  author={Zhang, Hongsheng and Huang, Yang},
  journal={Journal of High Energy Physics},
  volume={2021},
  number={12},
  pages={1--11},
  year={2021},
  publisher={Springer}
}

@article{huang2020scattering,
  title={Scattering of massless scalar field by charged dilatonic black holes},
  author={Huang, Yang and Zhang, Hongsheng},
  journal={The European Physical Journal C},
  volume={80},
  number={7},
  pages={654},
  year={2020},
  publisher={Springer}
}

@article{meng2023dynamics,
  title={Dynamics of null particles and shadow for general rotating black hole},
  author={Meng, Kun and Fan, Xi-Long and Li, Song and Han, Wen-Biao and Zhang, Hongsheng},
  journal={Journal of High Energy Physics},
  volume={2023},
  number={11},
  pages={1--28},
  year={2023},
  publisher={Springer}
}

@article{zhang2021poisson,
  title={Poisson-Arago spot for gravitational waves},
  author={Zhang, HongSheng and Fan, XiLong},
  journal={Science China Physics, Mechanics \& Astronomy},
  volume={64},
  number={12},
  pages={120462},
  year={2021},
  publisher={Springer}
}

@article{li2025gravitational,
  title={Gravitational Lensing of Gravitational Waves: Spin-wave Optics through Black Hole Scattering},
  author={Li, Zhao and Hou, Shaoqi and Zhao, Wen},
  journal={arXiv preprint arXiv:2512.23933},
  year={2025}
}

@article{li2025rigorous,
  title={Rigorous calculation of scalar scattering in the Schwarzschild background: The convergence of the partial-wave series and the Poisson spot},
  author={Li, Zhao and Zhao, Wen},
  journal={Physical Review D},
  volume={112},
  number={8},
  pages={083030},
  year={2025},
  publisher={APS}
}

@article{li2025absorption,
  title={Absorption and scattering of charged scalar waves by charged Horndeski black hole},
  author={Li, Qian and Wang, Qianchuan and Jia, Junji},
  journal={Physical Review D},
  volume={111},
  number={2},
  pages={024059},
  year={2025},
  publisher={APS}
}

@article{li2022absorption,
  title={Absorption cross section of regular black holes in scalar-tensor conformal gravity},
  author={Li, Yang and Miao, Yan-Gang},
  journal={Physical Review D},
  volume={105},
  number={4},
  pages={044031},
  year={2022},
  publisher={APS}
}

\end{document}